\documentclass [11pt]{article}
\usepackage{graphicx}
\usepackage{amsmath,amssymb} 
\usepackage{latexsym}
\usepackage{mathrsfs}
\usepackage{bbold}
\usepackage{color}
\usepackage{slashed}
\definecolor{red}{rgb}{1,0,0}
\usepackage{cancel}
\usepackage{comment}
\usepackage{cite}

\definecolor{red}{rgb}{1,0,0}

\makeatletter
\def\section{\@startsection {section}{1}{\z@}{-3.5ex plus -1ex minus
 -.2ex}{2.3ex plus .2ex}{\large\bf}}
\def\subsection{\@startsection{subsection}{2}{\z@}{-3.25ex plus -1ex
minus -.2ex}{1.5ex plus .2ex}{\normalsize\bf}}
\makeatother
\makeatletter

\@addtoreset{equation}{section}

\makeatother

\def\bea{\begin{eqnarray}} \def\eea{\end{eqnarray}}
\def\be{\begin{equation}} \def\ee{\end{equation}}

\newcommand{\promille}{%
  \relax\ifmmode\promillezeichen
        \else\leavevmode\(\mathsurround=0pt\promillezeichen\)\fi}
\newcommand{\promillezeichen}{%
  \kern-.05em%
  \raise.5ex\hbox{\the\scriptfont0 0}%
  \kern-.15em/\kern-.15em%
  \lower.25ex\hbox{\the\scriptfont0 00}}

\usepackage[colorlinks,linkcolor=black,citecolor=blue,urlcolor=blue,linktocpage]{hyperref}

\begin{document}

\thispagestyle{empty}

\begin{center}

\vspace*{-.6cm}

\hfill SISSA 40/2014/FISI \\

\begin{center}

\vspace*{1.1cm}

{\Large\bf  Bounds on OPE Coefficients in 4D Conformal Field Theories}
\end{center}

\vspace{0.8cm}

{\bf Francesco Caracciolo$^{a}$, Alejandro Castedo Echeverri$^{a}$,}\\
{ \bf Benedict von Harling$^{a}$ and Marco Serone$^{a,b}$}\\

\vspace{1.cm}

${}^a\!\!$
{\em SISSA and INFN, Via Bonomea 265, I-34136 Trieste, Italy} 

\vspace{.1cm}

${}^b\!\!$
{\em ICTP, Strada Costiera 11, I-34151 Trieste, Italy}

\end{center}

\vspace{1cm}

\centerline{\bf Abstract}
\vspace{2 mm}
\begin{quote}

We numerically study the crossing symmetry constraints in 4D CFTs, using previously introduced algorithms based on semidefinite programming.
We study bounds on OPE coefficients of tensor operators as a function of their scaling dimension and 
extend previous studies of bounds on OPE coefficients of conserved vector currents to the product groups SO$(N)\times$SO$(M)$.
We also analyze the bounds on the OPE coefficients of the conserved vector currents associated with the
groups SO$(N)$, SU$(N)$ and SO$(N)\times$SO$(M)$ under the assumption that in the singlet channel no scalar operator has dimension less than four, namely that the CFT has no relevant deformations.
This is motivated by applications in the context of composite Higgs models, where the strongly coupled sector is assumed to be a spontaneously broken 
CFT with a global symmetry.

\end{quote}


\vspace{1.1cm}

{\it To the memory of Francesco who tragically passed away during the completion of this project.}

\newpage

\tableofcontents

\section{Introduction}

There has recently been a renewed interest in studying general properties of four-dimensional (4D) Conformal Field Theories (CFTs) after the seminal paper \cite{Rattazzi:2008pe}
revived the bootstrap program advocated in the early 70s \cite{Ferrara:1973yt,Polyakov:1974gs}. Imposing the associativity of the Operator Product Expansion (OPE) 
and unitarity, ref.\cite{Rattazzi:2008pe} has shown
how one can set bounds on scalar operator dimensions in 4D CFTs. 
Although these constraints are based on numerical methods, they come from first principles, 
with no further assumptions.  
Since then, various  generalizations of this result have been developed in order to improve the above bounds, to put bounds on OPE coefficients and 
on CFT data in presence of a global symmetry \cite{Rychkov:2009ij, Caracciolo:2009bx,Poland:2010wg,Rattazzi:2010gj,Rattazzi:2010yc,Vichi:2011ux,Poland:2011ey}. In particular, bounds were derived on the OPE coefficients associated with the energy momentum tensor (the central charge) and the conserved 
vector current of a global symmetry \cite{Poland:2010wg,Rattazzi:2010gj,Rattazzi:2010yc,Vichi:2011ux,Poland:2011ey}.
Superconformal field theories and CFTs in $d\neq 4$ have also been considered \cite{Poland:2010wg,Poland:2011ey,ElShowk:2012ht,Beem:2013qxa,Gliozzi:2013ysa,Kos:2013tga,Gaiotto:2013nva,Berkooz:2014yda,El-Showk:2014dwa,Gliozzi:2014jsa,Nakayama:2014lva,Alday:2014qfa,Chester:2014fya}. In addition to the above mostly numerical results, considerable progress has also been made on more analytic aspects of CFTs, see e.g.
refs.\cite{Heemskerk:2009pn,Fitzpatrick:2010zm,Costa:2011mg,Costa:2011dw,SimmonsDuffin:2012uy,Pappadopulo:2012jk,Liendo:2012hy,Fitzpatrick:2012yx,Komargodski:2012ek,Hogervorst:2013sma,Fitzpatrick:2013sya,Hogervorst:2013kva,ElShowk:2011ag,Fitzpatrick:2011ia}.

The aim of this paper is to numerically study the bounds on the coefficient $\kappa$ of the two-point function between two conserved currents associated with a global symmetry of a CFT.
Our main motivation comes from theoretical considerations in the context of composite Higgs models, in which the CFT is the hidden sector which gives rise to the Higgs, and a subgroup of
the global symmetry of the CFT is weakly gauged in order to get the Standard Model gauge interactions.
These composite Higgs models models are related, through the AdS/CFT correspondence, to Randall-Sundrum theories \cite{Randall:1999ee} with
matter in the bulk, which are a promising solution to the gauge hierarchy problem. Particularly interesting are the models where the Higgs is a pseudo Nambu-Goldstone Boson (pNGB)
of an approximate spontaneously broken global symmetry of the CFT, which correspond to gauge-Higgs unification models in 5D warped theories.
Neither the UV completion of the 5D models nor the explicit form of the 4D CFT is known so far.
Calculability of the dual 5D models would require that the CFT is in some large N limit, but this is not a necessary requirement.
On the contrary, various phenomenological bounds tend to favour models at small N, so we will not assume the existence of a large N limit in the CFT.
Constructing such a CFT is not a trivial task, so we look for possible consistency relations. When the global symmetry of the CFT is gauged, the coefficient $\kappa$ of the current-current two-point function 
governs the leading contribution of the CFT to the one-loop evolution of the corresponding gauge coupling.  
This contribution should not lead to Landau poles  for the SM gauge couplings. We also require that the CFT has no relevant deformations, in order not to reintroduce the hierarchy problem.
This leads to the constraint that the dimension of the lowest-lying scalar singlet operator should be $\Delta_S\geq 4$. All our considerations apply independently of the pNGB nature or not of the Higgs. 

Motivated by the above considerations, we extend the analysis of ref.\cite{Poland:2011ey}, where lower bounds on $\kappa$ have been set 
starting from crossing constraints imposed on a four-point function of scalar operators in the fundamental representation of SO$(N)$ and SU$(N)$, in two ways.
First, we see how the bounds found in ref.\cite{Poland:2011ey} are modified when the lowest-lying singlet scalar operator is assumed to have a scaling dimension 
$\Delta_S\geq \Delta_{\rm min}$, where we choose $\Delta_{\rm min}=2,3,4$ for concreteness.
Second, we extend the analysis to non-simple groups of the form  SO$(N)\times$SO$(M)$.
We study non-simple groups because they easily allow to generalize the bounds for the groups SO$(N)$ and SU$(N)$, which are obtained by considering a single field
in the fundamental representation of the group, to multiple fields. Analogous to what was found in ref.\cite{Poland:2011ey} for singlet operators, the lower bounds on vector currents for SU$(N)$ groups that we find are, within the numerical precision, 
identical to those obtained for SO$(2N)$.
Hence we only report lower bounds for SO$(N)$ and SO$(N)\times$SO$(M)$ global symmetries.
We have derived the bootstrap equations also for groups of the form SO$(N)\times$SU$(M)$, but no bounds are reported for this case, since  
we have numerical evidence that the lower bounds for SO$(N)\times$SU$(M)$ are essentially identical to those
obtained for  SO$(N)\times$SO$(2M)$, similarly to the above equality between SO$(2N)$ and SU$(N)$ bounds.
In addition to that, we study the constraints on the OPE coefficients of spin $l=2$ and $l=4$ tensors coming from two identical scalar operators $\phi$, as a function of the scaling dimension of the tensors, 
in the general case in which no global symmetry is assumed. In analogy to the vector-current case, we analyze how these bounds change when one assumes a lower bound on the scale dimension of the scalar operators appearing in the $\phi\phi$ OPE.

All our numerical results are based on semi-definite programming methods, as introduced in ref.\cite{Poland:2011ey} in the context of the bootstrap approach, with a few technical modifications which are discussed in subsection \ref{sec:Num} and in appendix \ref{app:sdpa}.

The structure of the paper is as follows. In section 2 we describe the phenomenological motivations behind our work. In section 3 we briefly review the basic properties of 
the crossing constraints coming from four-point functions of identical scalars and review how bounds on OPE coefficients are numerically
obtained. In section 4 we report our results for the OPE coefficients of tensor $l=2$ and $l=4$ operators. Section 5 contains the most important results of the paper.
We report here the lower bounds on $\kappa$ associated with SO$(2N)$ (or SU$(N)$) vector currents, when the global symmetry of the CFT is SO$(2N)$ (or SU$(N)$)  and
$SO(2N)\times SO(M)$.
\footnote{Results for SO$(N)$ groups with odd $N$ are analogous to those for SO$(2N)$  and do not need any special treatment.}
In section 6 we conclude. 
Two appendices complete the paper. In appendix A we discuss various technical details about 
our implementation of the bootstrap equations in the semi-definite programming method, while in appendix B we report the crossing equations for SO$(N)\times$SO$(M)$ and SO$(N)\times$SU$(M)$.

\section{Motivation for a Gap in the Scalar Operator Dimension}

\label{sec:Pheno}

The motivation to consider CFTs with a gap in the scaling dimension of scalar gauge-singlet operators comes from
applications in the context of physics beyond the Standard Model (SM) that addresses the gauge hierarchy problem. 
The latter can be formulated from a CFT point of view, see e.g.~ref.\cite{Rattazzi:2008pe}.
Neglecting the cosmological constant,  the SM can be seen as an approximate CFT with one relevant deformation of classical mass dimension $\Delta_{H^\dagger H}=2$, corresponding to the Higgs mass term $H^\dagger H$. 
Relevant deformations grow in going from the UV towards the IR.
If we assume that the Higgs mass term is generated at some high scale $\Lambda_{UV}$, we would expect from naturalness
that the Higgs mass-squared term is of order $\Lambda_{UV}^{4-\Delta_{H^\dagger H}}=\Lambda_{UV}^2$ in the IR.
There are essentially two ways to solve this hierarchy problem: i) invoke additional symmetries that keep the relevant deformation small in the IR (e.g. supersymmetry);
ii) assume that the Higgs is a composite field of a strongly interacting sector, in which case the operator $H^\dagger H$ can have a large anomalous dimension that makes it
effectively marginal or irrelevant. 

A model along the lines of ii), conformal technicolor \cite{Luty:2004ye}, where the strongly coupled sector is assumed to be a CFT in the UV, 
was in fact the motivation for the pioneering work \cite{Rattazzi:2008pe}. 
Conformal technicolor is an interesting attempt to solve one of the long-standing problems of standard technicolor theories: how to reconcile
the top mass with Flavour Changing Neutral Current (FCNC) bounds.  In order to get a sizable top mass and at the same time avoid dangerous FCNCs, one has to demand that the scale dimension 
$\Delta_H$ of the Higgs field $H$ is as close to one as possible.  In order not to reintroduce the hierarchy problem, however, one has to keep $\Delta_{H^\dagger H}\gtrsim  4$ at the same time.
The analyses in refs.\cite{Rattazzi:2008pe,Poland:2011ey} have shown that generally these two conditions are in tension and that one needs $\Delta_H\gtrsim 1.52$ in order to have $\Delta_{H^\dagger H}\gtrsim 4$.

An alternative, phenomenologically more promising, solution is to rely on a different mechanism to generate SM fermion masses: partial compositeness \cite{Kaplan:1991dc}.
To this end,  one assumes that the SM fermions mix with fermion resonances of the strongly coupled sector. 
Due to this mixing, SM vectors and fermions become partially composite. In particular, the lighter the SM fermions are, the weaker is the mixing. This simple, yet remarkable, observation allows to significantly alleviate most flavor bounds. The Yukawa couplings are effective couplings that arise from the mixing terms once
the strongly coupled states are integrated out. 
 
This idea is particularly appealing when one assumes that the strongly coupled sector is an approximate CFT spontaneously broken at some scale $\mu$.
In this case, the hierarchy of the SM Yukawa couplings is naturally obtained by assigning different scale dimensions  $\Delta_\psi^i$ to the fermion operators mixing with the 
different SM fermions \cite{Agashe:2004rs}.  In particular, there is no longer 
the need to keep $\Delta_H$ close to one
since the effective size of the SM Yukawa couplings is governed by  $\Delta_\psi^i$. One assumes that $\Delta_\psi^i>5/2$ for all SM fermions except the top, so that the mixing terms are irrelevant 
deformations of the CFT and naturally give rise to suppressed Yukawa couplings in the IR. For the top, on the other hand, one assumes that $\Delta_\psi^t\simeq 5/2$, corresponding to a nearly marginal deformation of the CFT. 

One might wonder whether CFTs with all the necessary requirements to give rise to theoretically and phenomenologically viable composite Higgs models  exist at all.
A possible issue might arise in weakly gauging the  SM subgroup of the global symmetry of the CFT. Since partial compositeness requires a fermion operator in the CFT
for each SM fermion, dangerous Landau poles can potentially appear in the theory. 
Indeed, it has recently been shown that Landau poles represent the main obstruction in obtaining UV completions of composite Higgs models with a pNGB Higgs, based on supersymmetry \cite{Caracciolo:2012je}. 
It is then of primary importance to try to understand if and at what scale Landau poles will arise.
In theories with a pNGB Higgs, the relevant deformation $H^\dagger H$ can naturally be small, since it is protected by a shift symmetry. Moreover, it is not defined in the UV, where the global symmetry is restored. Nevertheless,  in order not to introduce other possible fine-tunings, one should demand that any scalar operator which is not protected by any symmetry, namely which is neutral under all possible global symmetries of the CFT, should be marginal or irrelevant. 

Summarizing, we can identify four properties that a CFT needs to have for a theoretically and phenomenologically viable composite Higgs model with partial compositeness:
\begin{enumerate}
\item{A global symmetry $G\supseteq G_{\rm SM}= \text{SU}(3)_c\times \text{SU}(2)_L\times \text{U}(1)_Y$.}
\item{No scalar operator with dimension $\Delta <4$ which is neutral under $G$.}
\item{No Landau poles for the SM gauge couplings below the scale $\Lambda_{\rm UV}$ when we gauge $G_{\rm SM}$.\footnote{Ideally, we might want to have $\Lambda_{\rm UV}\sim M_{\rm Planck}$.}}
\item{The presence of fermion operators with $\Delta_\psi^i\geq 5/2$ in some representation of $G$,  such that some of its components can mix with each of the SM fermion fields. 
At least one fermion operator should have dimension  $\simeq 5/2$.\footnote{The 
right-handed top, in principle, might be directly identified with a field of the CFT.}}
\end{enumerate}
Of course, these are only necessary but not sufficient conditions to get a viable CFT. 
In particular, one might want to address the mechanism which gives rise to the 
spontaneous breaking of the conformal symmetry as well as of the global symmetry in CFTs with a pNGB Higgs.

The consistency of a CFT which fulfils the above four conditions can be checked using crossing symmetry of four-point functions of the CFT. 
The first and second condition can be imposed by hand, assuming the existence of the global symmetry and that the lowest-dimensional scalar operator in the singlet channel has dimension $\Delta_S\geq 4$. One can extract information on the third condition by analyzing the bounds on the coefficients of current-current two-point functions.  
Finally, the fourth condition can again be implemented by assumption. The ideal configuration would be to analyze four-point functions involving fermion operators, which by assumption should appear in the CFT, and to extract any possible information from these correlators. 
Although this is in principle possible to do, correlation functions involving fermions in a non-supersymmetric setting have not been worked out so far. 
Postponing to a future project the analysis of such correlation functions,  in this paper we start to address these issues by replacing fermions with scalars with dimension $1\leq d < 2$ in the third requirement.

Let us estimate how severe the Landau pole problem can be in the simplest composite Higgs model where the Higgs is the pNGB associated with
the SO(5)$\rightarrow$SO$(4)$ symmetry breaking pattern. Let us consider the SU$(3)_c$ coupling $g_c$, 
because it runs fastest and possibly leads to the lowest-lying Landau pole, 
and let us denote by 
\be
\beta_{\rm CFT} = g_c ^3\frac{\kappa}{16\pi^2} 
\label{betaCFT0}
\ee
the CFT contribution to its one-loop $\beta$-function.  Assuming that the only non-SM fields which are charged under SU$(3)_c$ arise from the CFT,
a Landau pole develops at around
\be
\Lambda_L\simeq \mu \exp\bigg(\frac{2\pi}{(\kappa-7)\alpha_c(\mu)}\bigg)
\ee
for $\kappa>7$, where $\alpha_c=g^2_c/(4\pi)$ and $\mu\sim {\cal O}({\rm TeV})$ is the scale where the CFT breaks spontaneously.
Composite fermions coming from the CFT and mixing with SM fermions must be color triplets and in representations of SO$(5)$ that give rise to electroweak SU$(2)$ doublets and singlets.
If we assume them to be  in the fundamental representation ${\bf 5}$ of SO$(5)$, the fermion components in a given ${\bf 5}$ can mix with both the 
left-handed and right-handed components of a quark field.
We then need $n_f=6$ ${\bf 5}$s, one for each quark field, for a total of $6\times 5 = 30$ SU$(3)_c$ triplet Dirac fermions. In order to have an idea of the scales which are involved, it is useful
to consider the (unrealistic) limit of a free CFT. In this case, we get
\be
\kappa_{\rm free} = \frac 23 \times 30 = 20\,,
\label{kSO5free}
\ee
corresponding to  
\be
\Lambda_L\sim 200\;\; {\rm TeV} \,,
\label{Landau}
\ee
for $\mu\simeq 1$ TeV. It is clearly very important to set lower bounds on $\kappa$ in a generic CFT, given the exponential sensitivity of $\Lambda_L$ on this quantity. 

In the following, we will analyze bounds on the coefficients for SO$(2N)$ (or, equivalently, SU$(N)$) currents obtained from four-point functions of scalar operators in the fundamental representation of the group in presence of a gap in the operator dimension in the scalar gauge-singlet channel. In order to mimic the presence of more than one field multiplet, we will also consider fields in the bi-fundamental representation of the product group SO$(2N)\times$SO$(M)$.

\section{Review of the Bootstrap Program}

\label{sec:bootstrap}

In this section, we briefly review the equations that one obtains by imposing crossing symmetry on four-point functions of scalar operators in a unitary CFT, and how
these are numerically handled to get bounds on the CFT data. We refer the reader to ref.\cite{Rattazzi:2008pe} and references therein for more details and background material.

\subsection{Bootstrap Equations}

The four-point function of identical real scalar operators with scale dimension $d$ in a 4D CFT can be written as
 \be
 \langle \phi(x_1)\phi(x_2)\phi(x_3)\phi(x_4)\rangle = \frac{g(u,v)}{x_{12}^{2d}x_{34}^{2d}}\,,
 \label{4ptfunc}
 \ee 
where $x_{ij}^2=(x_i-x_j)_\mu (x_i-x_j)^\mu$ and
\be
u=\frac{x_{12}^2x_{34}^2}{x_{13}^2x_{24}^2}\,, \ \ v = \frac{x_{14}^2x_{23}^2}{x_{13}^2x_{24}^2}
\label{uvvariabl}
\ee
are two conformally invariant variables. All the dynamics is encoded in the function $g(u,v)$.
Using the OPE between $\phi(x_1) \phi(x_2)$ and $\phi(x_3) \phi(x_4)$ (the $s$-channel), in the region $0\leq u,v \leq 1$ this function is found to be
\be
g(u,v) = 1+\sum_{\Delta,l} |\lambda_{\phi\phi{\cal O}}|^2 g_{\Delta,l}(u,v)\,,
\label{confblocks}
\ee
where the sum is over all primary (traceless symmetric) operators of dimension $\Delta$ and (even) spin $l$ that appear in the OPE and $\lambda_{\phi\phi{\cal O}}$ is the 
coefficient of the three-point function $\langle \phi \phi {\cal O}\rangle$. 
The $+1$ in eq.~(\ref{confblocks}) results from the contribution of the identity operator which is present in the OPE of two identical operators. The three-point function $\langle \phi \phi {\cal O}\rangle$ in this case simplifies to the
two-point function $\langle \phi \phi \rangle $  which is normalized to unity. For each primary operator ${\cal O}$, the function $g_{\Delta,l}(u,v)$, which is called a conformal block, takes into account the contribution of all descendants of ${\cal O}$.
In 4D, the explicit form of  $g_{\Delta,l}(u,v)$ is known and reads \cite{Dolan:2000ut,Dolan:2003hv}
\be\begin{split}
g_{\Delta,l}(u,v) & = \frac{z\bar z}{z-\bar z}\Big(k_{\Delta+l}(z) k_{\Delta-l-2}(\bar z) - (z\leftrightarrow \bar z) \Big)\,, \\
k_\beta(x) & \equiv x^{\beta/2} {}_2F_1\Big(\frac\beta2,\frac\beta2,\beta,x\Big)\,,
\label{conblockExp}
\end{split}\ee
where $u = z\bar z$ and $v=(1-z)(1-\bar z)$.\footnote{Here we have used the normalization of the conformal blocks introduced in ref.\cite{Rattazzi:2010yc} which differs by a factor $(-2)^l$ from the 
one of refs.\cite{Dolan:2000ut,Dolan:2003hv}.}
Alternatively, we can obtain an expression for $g(u,v)$ using the OPE between $\phi(x_2) \phi(x_3)$ and $\phi(x_1) \phi(x_4)$ (the $t$-channel). Demanding that the $s$-channel and $t$-channel results for the four-point function
agree gives the crossing symmetry constraint (or bootstrap equation)
\be
 \sum_{\Delta,l}  |\lambda_{{\cal O}}|^2 F_{d,\Delta,l}(z,\bar z) = 1\,, 
\label{BootEQ} 
\ee
with
\be
F_{d,\Delta,l}(z,\bar z) \equiv \frac{v^d g_{\Delta,l}(u,v) - u^d g_{\Delta,l}(v,u)}{u^d-v^d}
\label{FDef}
\ee
and $\lambda_{{\cal O}} \equiv \lambda_{\phi\phi{\cal O}}$.
When the CFT has a global symmetry $G$, the above analysis can be generalized using scalar fields $\phi_a$ (real or complex) in some representation $r$ of $G$ \cite{Rattazzi:2010yc}.
The symmetry implies that all the field components of the multiplet must have the same dimension $d$. Moreover, it allows to easily classify the
operators appearing in the $\phi_a\phi_b$ OPE in terms of the irreducible representations appearing in the product $r\otimes r$. 
A similar analysis applies for complex fields in the $\phi_a\phi_b^\dagger$ OPE. 
It is useful to introduce another function, similar to the $F$ of eq.~(\ref{FDef}):
\be
H_{d,\Delta,l}(z,\bar z) \equiv \frac{v^d g_{\Delta,l}(u,v) + u^d g_{\Delta,l}(v,u)}{u^d+v^d}\,.
 \label{HDef}
\ee
In presence of a global symmetry $G$, eq.~(\ref{BootEQ}) generalizes to a system of $P+Q$ equations of the form
\be\begin{split}
& \sum_{i} \eta_{F,i}^p  \sum_{{\cal O}\in r_i} |\lambda_{{\cal O}_i}|^2 F_{d,\Delta,l}(z,\bar z) = \omega_F^p\,, \ \ p=1,\ldots,P\,, \\
& \sum_{i} \eta_{H,i}^q  \sum_{{\cal O}\in r_i} |\lambda_{{\cal O}_i}|^2 H_{d,\Delta,l}(z,\bar z) = \omega_H^q\,, \ \ q=P+1,\ldots, P+Q\,.
\label{GenBE}
\end{split}\ee
Here, $i$ runs over all possible irreducible representations that can appear in the $s$- and $t$-channel decomposition, $\eta_{F,i}^p$ and $\eta_{H,i}^q$ are numerical factors that depend on $G$ and 
$\lambda_{{\cal O}_i}$ is a short-hand notation for the $\langle \phi_a\phi_b {\cal O}\rangle$ three-point function coefficient. Furthermore,
$\omega_F^p=1$ and $\omega_H^q=-1$ if the singlet representation appears in the left-hand side of eq.~(\ref{GenBE}), and $\omega_F^p= \omega_H^q=0$ otherwise.
The explicit form of eq.~(\ref{GenBE}) for the cases of interest will be given in section \ref{sec:global}  and appendix \ref{app:crossing}.

\subsection{Bounds on OPE Coefficients and Numerical Implementation}

\label{sec:Num}

The bootstrap equation (\ref{BootEQ}) has originally been used to set bounds on the  scalar operator dimensions that can appear in a CFT.
Shortly after that, ref.\cite{Caracciolo:2009bx} has shown how to obtain bounds on the OPE coefficient $\lambda_{{\cal O}_0}$ of an operator ${\cal O}_0$ appearing in the $\phi \phi$ OPE.
Let us assume that a linear functional $\alpha$ can be found, such that
\be
\alpha(F_{d,\Delta_0,l_0}) = 1\,, \ \ \ \ \ \ \alpha(F_{d,\Delta,l}) \geq 0   \ \ \ \  \forall (\Delta,l)\neq (\Delta_0,l_0)\,.
\label{alpha}\ee
Applying such a functional to eq.~(\ref{BootEQ}) gives
\be
|\lambda_{{\cal O}_0}|^2 = \alpha(1) - \hspace*{-.3cm} \sum_{(\Delta,l)\neq (\Delta_0,l_0)} |\lambda_{{\cal O}}|^2 \alpha(F_{d,\Delta,l}) \, \leq \, \alpha(1)\,.
\label{boundequation}
\ee
The optimal bound is obtained by minimizing $\alpha(1)$ among all the functionals $\alpha$ which satisfy eq.~(\ref{alpha}).
One can use the functional $\alpha$ also to rule out the existence of certain CFTs. 
For instance, if under a certain assumption on the CFT data one finds a functional $\alpha$ and an operator ${\cal O}_0$ for which $|\lambda_{{\cal O}_0}|^2<0$,
then that CFT is ruled out.

The above procedure is easily generalized in presence of global symmetries. Let us assume that we want to bound the OPE coefficient of an operator ${\cal O}_0$ with dimension
$\Delta_0$ and spin $l_0$ in the representation $r_1$. We look for a set of linear functionals $\alpha_m$ ($m=1,\ldots,P+Q$) such that
\be\begin{split}
& \sum_{p=1}^P \alpha_p\Big(\eta_{F,1}^p F_{d,\Delta_0,l_0} \Big)+\sum_{q=P+1}^{P+Q} \alpha_q\Big(\eta_{H,1}^q H_{d,\Delta_0,l_0}\Big) = 1\,, \\
& \sum_{p=1}^P \alpha_p\Big(\eta_{F,1}^p F_{d,\Delta,l}\Big)+\sum_{q=P+1}^{P+Q} \alpha_q\Big(\eta_{H,1}^q H_{d,\Delta,l}\Big) \geq 0\,, \ \ \  \ \ \forall (\Delta,l)\neq (\Delta_0,l_0)\,, \\
& \sum_{p=1}^P \alpha_p\Big(\eta_{F,i}^p F_{d,\Delta,l}\Big)+\sum_{q=P+1}^{P+Q} \alpha_q\Big(\eta_{H,i}^q H_{d,\Delta,l}\Big) \geq 0 \,,\ \ \ \ \ \forall (\Delta,l) \, , i\neq 1\,.
\label{alphaVect}
\end{split}\ee
Applying such a functional to eq.~(\ref{GenBE}) gives 
\be
|\lambda_{{\cal O}_0}|^2\leq  \sum_{p=1}^P \alpha_p(\omega_p^F)+\sum_{q=P+1}^{P+Q} \alpha_q(\omega_q^H)\,.
\ee
In our paper, we will mainly be interested in the OPE coefficient associated with a conserved vector current $J_\mu$ of a global symmetry, which has  $\Delta_0=3$ and $l_0=1$.
We shall denote this coefficient by $\lambda_J$.  As we will discuss in section \ref{sec:global} (see eq.~(\ref{lambdaKappa})), upper bounds on $|\lambda_J|^2$ turn into
lower bounds on the coefficient $\kappa$ introduced in eq.~(\ref{betaCFT0}). 

Following ref.\cite{Rattazzi:2008pe}, 
we consider functionals that act as linear combinations of derivatives on a generic function $f(z,\bar z)$,
\be
\alpha(f(z,\bar z)) = \sum_{m+n\leq 2k} a_{mn} \partial_z^m\partial_{\bar z}^n f(z,\bar z)|_{z=\bar z = 1/2}\,,
\label{alphaExp}
\ee
where $a_{mn}$ are real coefficients. Due to the symmetries of the conformal blocks $F$ and $H$, the sum can be restricted to $m<n$ and even values of $m+n$ when $\alpha$ acts on $F$,
and $m<n$ and odd values of $m+n$ when it acts on $H$. 

We numerically search for functionals $\alpha$ which satisfy eqs.~(\ref{alpha}) and (\ref{alphaVect}) by following the method developed in refs.\cite{Poland:2010wg,Poland:2011ey}.
We refer the reader to  these references for further details. For this method, one approximates the derivatives of the conformal blocks $F_{d,\Delta,l}$ and $H_{d,\Delta,l}$  in eq.~(\ref{alphaExp})
with polynomials $P^{mn}_l(\Delta_l(1+x))$, where $x\in [0,\infty)$ and $\Delta_l = l+2$ 
is the unitarity bound on the scaling dimension for an operator of spin $l$ ($\Delta_0=1$ for $l=0$).
The requirements in eq.~(\ref{alpha}) or eq.~(\ref{alphaVect}) imply that the linear combination of polynomials of the form $a_{mn}P^{mn}_l$ must be positive-semidefinite on the positive real $x$-axis, for any value of $l$. 
There are two great virtues in setting up the problem in this way. Firstly, there is no need to discretize the dimension $\Delta$ and to put a cut-off value $\Delta_{\rm max}$, like in the linear programming methods used in ref.\cite{Rattazzi:2008pe}. In particular, we can probe all $\Delta$ continuously up to infinity. Secondly, one can exploit numerical packages that allow to handle very large systems of equations quite efficiently.
A key variable in the numerical algorithm is the coefficient $k$ entering in eq.~(\ref{alphaExp}). The larger $k$, the larger is the space of possible viable functionals, and hence the stronger are the bounds.
Of course, the larger $k$, the more time-consuming is the numerical evaluation. For our computations, we have chosen $k=9,10,11$, depending on the complication of the problem. 

The above algorithm, however, still requires to truncate the system at a given maximal spin $L$. This is in principle a serious problem, because one might have 
\be
\alpha(F_{d,\Delta,l})< 0 \ \ \ \ \ {\rm for} \ \  l>L\,.
\label{lgelMax}
\ee
If $L$ is chosen sufficiently large, ${\cal O}(10)$ or more, we do not expect possible violations in the semidefinite positiveness of $\alpha$ of the form (\ref{lgelMax}) to be important for the numerical value of the bound. Indeed,  large spin $l$ implies large dimensions $\Delta$ according to the unitarity bound, and the contribution to the four-point function of operators with large $\Delta$ is exponentially suppressed in $\Delta$ \cite{Pappadopulo:2012jk}.
Nevertheless, it would be more reassuring to have more control on such effects.
For parametrically large $l$, the conformal blocks $F_{d,\Delta,l}$ and  $H_{d,\Delta,l}$ and their derivatives allow for simple analytic expressions.
For large $l$, the terms involving the highest derivatives dominate.
Using these analytic expressions, we can find the value $l_{\rm max}$, which {\it depends on $k$}, for which the contribution of the large-$l$ conformal blocks is largest.
We find, for $2k\gg 1$ (see appendix \ref{app:sdpa} for details)
\be
l_{\rm max} \sim  \frac{2k}c\,,
\label{lD}\ee
where $c=-\log(12-8\sqrt{2})\simeq 0.377$. For $k\sim 10$, eq.~(\ref{lD}) gives $l_{\rm max}\sim 50\div 60$.  Ideally, one would include all spins from $l=0$ up to $L=l_{\rm max}$.
This is computationally quite demanding. Fortunately, we have found that it is sufficient to take $L=20$ to get numerically stable bounds.
Changing $L$ to  $L=22$ or $L=24$ does not significantly alter the bounds. Nevertheless, in order to have more control on the higher-$l$ states, 
we have included two other states in the constraints, at $l=l_{\rm max}$ and at an intermediate value $l\approx (L+l_{\rm max})/2$.\footnote{More precisely, we include spins $l=35,52$ for calculations with $k=9$, $l=37,56$ for $k=10$ and $l=40,60$ for $k=11$.}
We have numerically tested that this implementation works better than including states at very large values of $l$, such as $l=1000, 1001$ as done in e.g. ref.\cite{Poland:2011ey}. We can always check the positivity of $\alpha$ a posteriori. We have found that by imposing constraints at $l = 0,...,20,1000,1001$
the functional often becomes negative for values $l \neq 0,...,20,1000,1001$ whereas for our implementation $\alpha$ remains positive for most of the $l$ that we have checked.
In practice, however, we have not detected deviations in the results among the two different implementations, confirming that values of $l>L$ are numerically negligible.

\begin{figure}[!t]
\begin{center}
\fbox{l=0 operators} \\[0.05cm]
		\includegraphics[width=110mm]{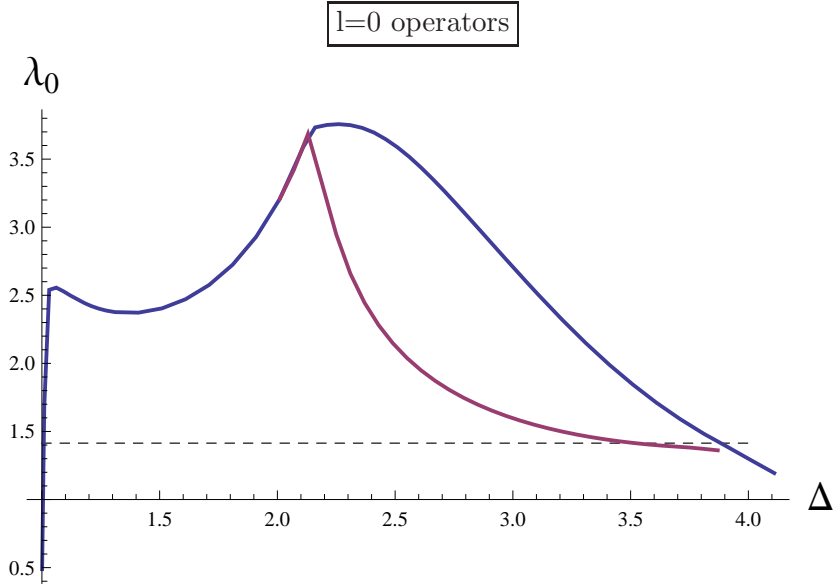}
\end{center}
\caption{\label{fig2} 
\small
Upper bounds on the three-point function coefficient $\lambda_0$ between two scalar operators of dimension $d=1.6$ and a scalar operator ${\cal O}$ of dimension
$\Delta$ calculated at $k=11$ with no assumptions on the spectrum (blue line) and assuming that no scalar operator in the OPE is present below $\Delta_0 = 2$ (red line). 
For illustrative purposes, we show the free-theory value for $d=1$ (in which case $\Delta = 2$), $\lambda^{\rm free}_0 = \sqrt{2}$, as a black dashed line.}
\label{fig:scalarGap2}
\end{figure}

\section{Bounds on OPE Coefficients for Tensor Operators}
\label{BoundsNoS}

In this section, we report our results for the upper bounds on the three-point function coefficient $\lambda_{{\cal O}}$ 
appearing in the OPE of  two identical scalar operators $\phi$ of scaling dimension $d$.
The operator ${\cal O}$ is a traceless symmetric tensor of even spin $l$.  
The coefficient $\lambda$ is normalized such that its free-theory value is 
\be
\lambda_{{\cal O}_l}^{\rm free} \equiv \lambda_l^{\rm free} = \sqrt{2}\frac{l!}{\sqrt{(2l)!}}\,.
\label{Scalar1}
\ee
We do not report the results for the $l=0$ case, which were first derived in ref.\cite{Caracciolo:2009bx} and subsequently improved in ref.\cite{Poland:2011ey}. Our results agree with fig.~10 of ref.\cite{Poland:2011ey}.
These bounds change if we assume that the first scalar operator which appears in the $\phi \phi$ OPE has a dimension $\Delta_0 > \Delta_l$, where $\Delta_l$ is the unitarity bound on $\Delta$.
As expected, the upper bounds do not significantly change when $d$ is close to 1, since by continuity the theory is close to the free theory, 
where the only scalar operator arises exactly at $\Delta_0=2$.
For values of $d$ not too close to 1, on the other hand,  the bound is significantly improved and becomes more stringent as $\Delta_0$ increases.
In fig.~\ref{fig:scalarGap2}, we report the bounds for $d=1.6$ and $\Delta_0=2$. 

\begin{figure}[!t]
\begin{center}
\hspace*{-0.65cm} 
\begin{minipage}{0.5\linewidth}
\begin{center}
	\hspace*{0cm} 
	\fbox{\footnotesize l=2 operators} \\[0.05cm]
	\includegraphics[width=70mm]{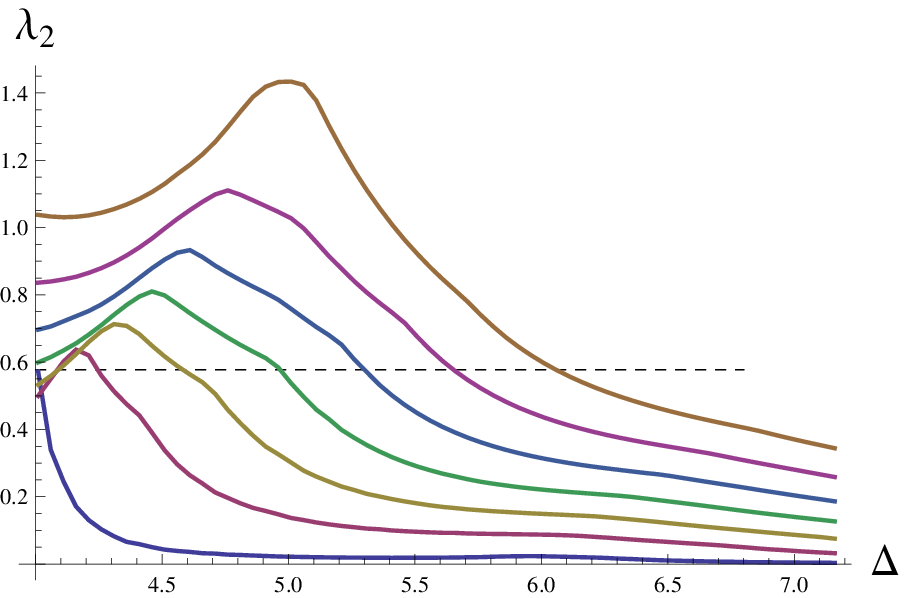}\\
	\mbox{\footnotesize (a)} \\
\end{center}
\end{minipage}
\hspace{0.25cm}
\begin{minipage}{0.5\linewidth}
\begin{center}
	\hspace*{0cm} 
	\fbox{\footnotesize l=2 operators} \\[0.05cm]
	\includegraphics[width=70mm]{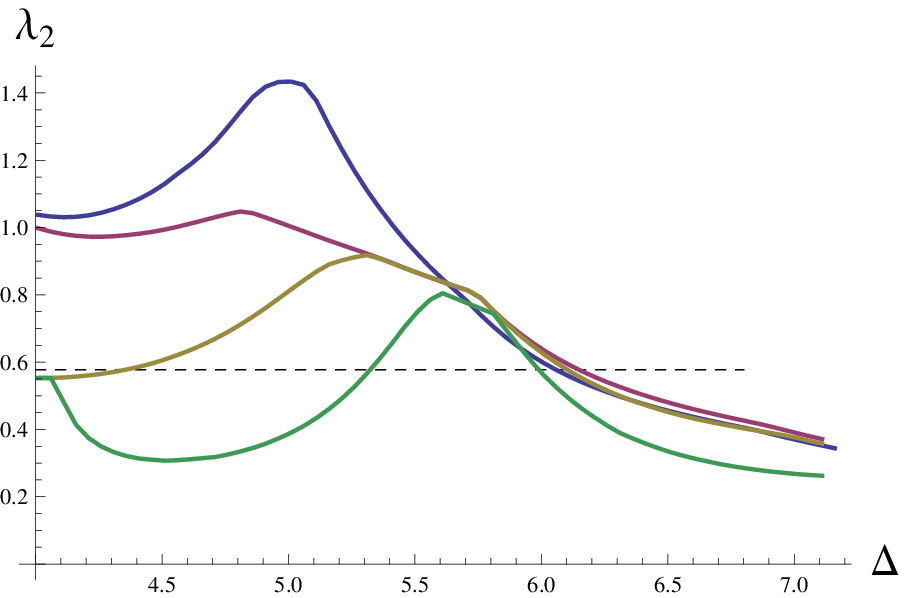}\\
	\mbox{\footnotesize (b)} \\
\end{center}
\end{minipage}
\end{center}
\vspace*{-0.2cm}
\caption{\label{fig:tensorNogapGap2} 
\small
Upper bounds on the three-point function coefficient $\lambda_2$ between two scalar operators of dimension $d$ and a tensor operator ${\cal O}$ with spin $l=2$ and dimension
$\Delta$ calculated at $k=11$. (a) Starting from below, the lines correspond to the values $d=1.01, 1.1,1.2,1.3,1.4,1.5,1.6$. No assumption on the spectrum is made. (b) For $d=1.62$ with no assumption on the spectrum (blue line)  and assuming that no scalar operator in the OPE is present below $\Delta_0 = 2$ (red line), $\Delta_0 = 3$ (brown line) and $\Delta_0 = 4$ (green line).
For illustrative purposes, we show the free-theory value for $d=1$ (in which case $\Delta = 4$), $\lambda^{\rm free}_2 = 1/\sqrt{3}$, as a black dashed line in both panels.}
\end{figure}
\begin{figure}[!t]
\begin{center}
\hspace*{-0.65cm} 
\begin{minipage}{0.5\linewidth}
\begin{center}
	\hspace*{0cm} 
	\fbox{\footnotesize l=4 operators} \\[0.05cm]
	\includegraphics[width=70mm]{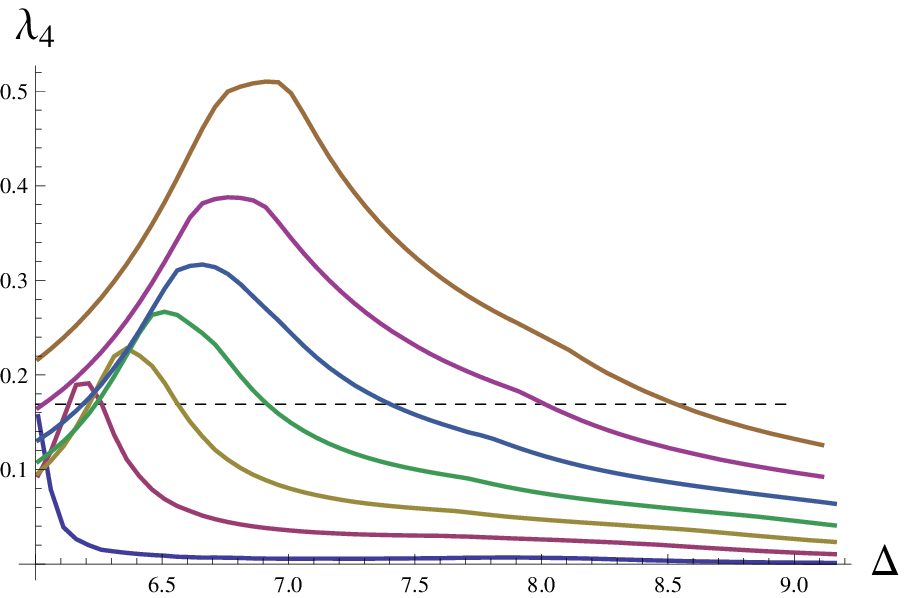}\\
	\mbox{\footnotesize (a)} \\
\end{center}
\end{minipage}
\hspace{0.25cm}
\begin{minipage}{0.5\linewidth}
\begin{center}
	\hspace*{0cm} 
	\fbox{\footnotesize l=4 operators} \\[0.05cm]
	\includegraphics[width=70mm]{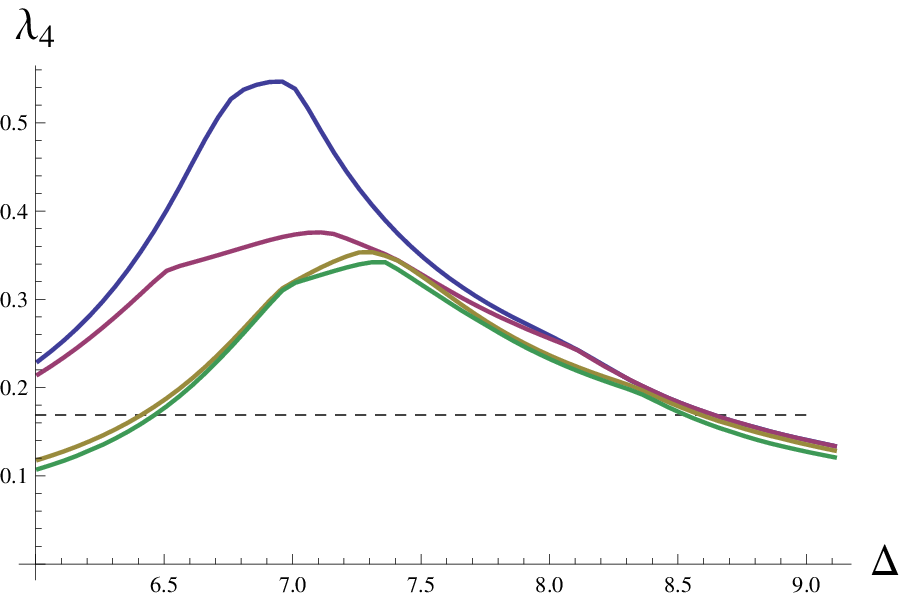}\\
	\mbox{\footnotesize (b)} \\
\end{center}
\end{minipage}
\end{center}
\vspace*{-0.2cm}
\caption{\label{fig:4tensorNogapGap2} 
\small
Upper bounds on the three-point function coefficient $\lambda_4$ between two scalar operators of dimension $d$ and a tensor operator ${\cal O}$ with spin $l=4$ and dimension
$\Delta$ calculated at $k=11$. (a) Starting from below, the lines correspond to the values $d=1.01, 1.1,1.2,1.3,1.4,1.5,1.6$. No assumption on the spectrum is made. (b) For $d=1.62$ with no assumption on the spectrum (blue line)  and assuming that no scalar operator in the OPE is present below $\Delta_0 = 2$ (red line), $\Delta_0 = 3$ (brown line) and $\Delta_0 = 4$ (green line).
For illustrative purposes, we show the free-theory value for $d=1$ (in which case $\Delta = 6$), $\lambda^{\rm free}_4 = 1/\sqrt{35}$, as a black dashed line in both panels.}
\end{figure}

Analogously, one can study the upper bounds on $\lambda_2$ for generic tensor operators ${\cal O}$ with spin $l=2$ and dimension $\Delta\geq 4$. Upper bounds on the central charge $c\propto 1/\lambda_2^2$ 
associated with the energy-momentum tensor (the lowest-dimensional operator in the $l=2$ sector, with $\Delta=4$), have been extensively analyzed in refs.\cite{Poland:2010wg,Rattazzi:2010gj,Poland:2011ey},
with and without the assumption of a lower bound on the dimension of the lowest-lying scalar operator appearing in the $\phi  \phi$ OPE.
In fig.~\ref{fig:tensorNogapGap2} (a), we report the upper bounds on the coupling $\lambda_2$  between two scalar operators of dimension $d$ and a tensor operator ${\cal O}$ with spin $l=2$ and dimension $\Delta$ for different 
$d$ and as a function of $\Delta$.
As can be seen, the larger $d$ is, the less stringent is the upper bound, in agreement with the naive expectation for 
which $d-1$ can be seen as a measure (for $d$ not too far from 1) of how strongly coupled the CFT is. 
Like for scalar operators, the bounds change if we make some assumptions on the CFT spectrum. 
As for the scalar case, the upper bounds do not significantly change when $d$ is very close to 1, but for values of $d$ not too close to 1, they become more stringent as $\Delta_0$ increases.
For illustration, in fig.~\ref{fig:tensorNogapGap2} (b), we report the upper bounds on $\lambda_2$ as a function of $\Delta$ for $d=1.62$,  assuming that the lowest {\it scalar} operator appearing in the $\phi \phi$ OPE has a dimension $\Delta_0\geq 2$,  $\Delta_0\geq 3$ and $\Delta_0\geq 4$.\footnote{The value $d\simeq 1.62$ is roughly the minimal one compatible with the assumption $\Delta_0\geq 4$,  see e.g. fig.~2 of ref.\cite{Poland:2011ey}.}

Similarly, one can analyze tensor operators at higher $l$. In figs.~\ref{fig:4tensorNogapGap2} (a) and (b), we report the same as above for $l=4$ operators.
As expected, the absolute scale of $\lambda_l$ becomes lower and lower as $l$ increases, with the allowed values of $\lambda_l$ quickly decreasing as $l$ becomes larger.
Notice that the maximal allowed value of both $\lambda_2$ and $\lambda_4$ is centered at  values of $\Delta$ that increase as $d$ is increased.

\section{Bounds on Current-Current Two-Point Functions}

\label{sec:global}

At leading order, the CFT contribution to the one-loop beta function of a gauge field $A_\mu$, external to the CFT,  
is governed by the coefficient of the two point-function of the corresponding current. Denoting by 
\be
{\cal L}_{\rm gauged} = {\cal L}_{\rm CFT}+ g J^\mu_A A_\mu^A -\frac 14 F_{\mu\nu}^AF^{\mu\nu}_A
\ee
the total Lagrangian after the gauging, we can consider the effective action  $\Gamma(A)$ defined as (in euclidean signature)
\be
e^{-\Gamma(A)} = \int{\cal D}\Phi_{\rm CFT} \, \,  e^{-\int d^4 x \, {\cal L}_{\rm gauged}}\,,
\label{EffActionAmu}
\ee
where the functional integration is over all the CFT states and we have omitted color indices.
In general
\be
\Gamma(A) \supset -\frac 14 \int d^4x \, Z F_{\mu\nu}^AF^{\mu\nu}_A \,,
\ee
where $Z=(1+\delta Z_{\rm CFT})$ and $\delta Z_{\rm CFT}$ is the CFT contribution to the wave function renormalization of the gauge field, which in turns
gives us the one-loop contribution of the CFT to the RG running of $g$:
\be
\beta_{\rm CFT} = g \mu\frac{d}{d\mu} \sqrt{Z} = \frac 12 g \mu\frac{d}{d\mu} \delta Z_{\rm CFT} \,.
\ee
By taking two functional derivatives with respect to $A^A_\mu(p)$ and $A_\nu^B(-p)$ in eq.~(\ref{EffActionAmu}), we readily get 
\be
\delta_{AB} \delta Z_{\rm CFT} (\delta_{\mu\nu} p^2 - p_\mu p_\nu) = -g^2 \langle J_\mu^A(-p) J_\nu^B(p) \rangle_{g=0} \,,
\ee
where the subscript in the correlator specifies that the two-point function is computed in the {\it unperturbed} CFT setting $g=0$.
The normalization of the current is uniquely fixed by Ward identities. Following the notation of ref.\cite{Poland:2011ey}, we parametrize the two-point function in configuration space as follows:\footnote{Notice that
the definition of $\kappa$ here is not identical to that of ref.\cite{Poland:2011ey} which tacitly applies to CFTs with one charged multiplet only. In general, $\kappa_{\rm here} \propto \sum_i \kappa^i_{\rm there} T(r_i)$ where $i$ runs over all the charged fields of the CFT in the representations $r_i$ and $\delta^{AB} T(r_i)={\rm Tr}(t^A_{r_i} t^B_{r_i})$.\label{kappafootnote}} 
\be
\langle J^A_\mu(x) J^B_\nu(0) \rangle_{g=0} = \frac{3 \kappa \delta^{AB}}{4\pi^4} \Big(\delta_{\mu\nu}-2 \frac{x_\mu x_\nu}{x^2}\Big) \frac{1}{x^6} \,.
\label{kappadef}
\ee
The ``vector central charge" $\kappa$  is roughly a measure of how many charged degrees of freedom are present in the CFT, similar to the standard central charge $c$ being a
measure of the total number of degrees of freedom of the CFT.
Modulo irrelevant contact terms, the momentum space correlation function reads 
\be
\langle J_\mu^A(-p) J_\nu^B(p) \rangle_{g=0}  = (\delta_{\mu\nu} p^2 - p_\mu p_\nu)  \frac{\kappa}{16\pi^2} \delta^{AB} \log\Big(\frac{p^2}{\mu^2}\Big)
\ee
and hence
\be
\beta_{\rm CFT} = g ^3\frac{\kappa}{16\pi^2} \,.
\label{betaCFT}
\ee
We extract $\kappa$ by rescaling the vector current so that it appears as the coefficient of the three-point function $\langle \phi_i \phi_j J_\mu^A\rangle$:
\be
\lambda_J^2 = \frac{\rho}{\kappa}\,.
\label{lambdaKappa}
\ee
Upper bounds on $\lambda_J^2$ turn into lower bounds on $\kappa$.
The constant factor $\rho$ is easily found by matching the result with the free-theory case, in which both $\lambda_J^2$ and $\kappa$ are calculable.
In what follows, we will analyze the lower bounds on $\kappa$ for different vector currents that come from the crossing symmetry constraints applied to four-point functions of scalars.

\subsection{SO$(N)$ Global Symmetry}
\label{sec:soN}

We consider a four-point function of real scalars that are taken to be the components of a single field in the fundamental representation of SO$(N)$ with dimension $d$.
The crossing symmetry relations have been derived in ref.\cite{Rattazzi:2010yc}. We report them here for completeness:
\be
\sum_{S^+} |\lambda_{\cal O}^S|^2 \left(\begin{array}{c}
0\\
F \\
H \end{array}\right)
+\sum_{T^+} |\lambda_{\cal O}^T|^2 \left(\begin{array}{c}
F \\
(1-\frac 2N)F \\
-(1+\frac 2N)H \end{array}\right)
+\sum_{A^-} |\lambda_{\cal O}^A|^2 \left(\begin{array}{c}
-F \\
F \\
-H \end{array}\right) = 
 \left(\begin{array}{c}
0\\
1 \\
-1 \end{array}\right).
\label{CrossingSON}
\ee
Here $S$, $T$ and $A$ refer respectively to the singlet, rank-2 symmetric and antisymmetric (adjoint) representations of the operators ${\cal O}$ which define the 
different conformal blocks. 
For the superscript $+$, only even spins are included in the sum whereas for $-$ only odd spins are summed over.
For simplicity, we have omitted
the labels $d,\Delta,l$ and the arguments $z,\bar z$ of the conformal blocks $F$ and $H$. 
Bounds on $\kappa$ (as defined in eq.~\eqref{kappadef}) in this set-up  have already been found in ref.\cite{Poland:2011ey}. In this subsection we will see how these bounds change if assumptions on the 
dimensionality  of the lowest-lying scalar operator in the singlet channel are made.

First of all, let us consider the free theory of a real scalar in the fundamental representation of SO$(N)$ in order to fix the constant $\rho$ in eq.~(\ref{lambdaKappa}).
The free-theory values of the OPE coefficients in the three different channels read
\be
\begin{split}
\lambda_{A,l}^{\rm free} & =\frac{1}{\sqrt{2}} \lambda_l^{\rm free}  \ \ \ \ \ \ (l\; {\rm odd})\,, \\ 
\lambda_{T,l}^{\rm free} & = \frac{1}{\sqrt{2}} \lambda_l^{\rm free}  \ \ \ \ \ \ (l\; {\rm even})\,, \\
\lambda_{S,l}^{\rm free} & = \frac 1{\sqrt{N}} \lambda_l^{\rm free} \ \ \ (l\; {\rm even}) \,,
\end{split}\ee
where $\lambda_l^{\rm free}$ is given in eq.~(\ref{Scalar1}). 
We in particular get  $\lambda_J^{\rm free}=\lambda_{A,1}^{\rm free} = 1/\sqrt{2}$.  Matching eq.~(\ref{betaCFT}) with the one-loop contribution to the $\beta$-function of a scalar in an SO$(N)$ gauge theory gives
\be
\kappa_{\rm free}  =\frac 16 \,,
\label{KfreeSON}
\ee
where we have taken $T({\rm fund.}) = 1$ (cf.~footnote \ref{kappafootnote}).  
From this it follows that $\rho=1/12$ in eq.~(\ref{lambdaKappa}).

\begin{figure}[!t]
\begin{center}
\hspace*{0cm} 
\includegraphics[width=110mm]{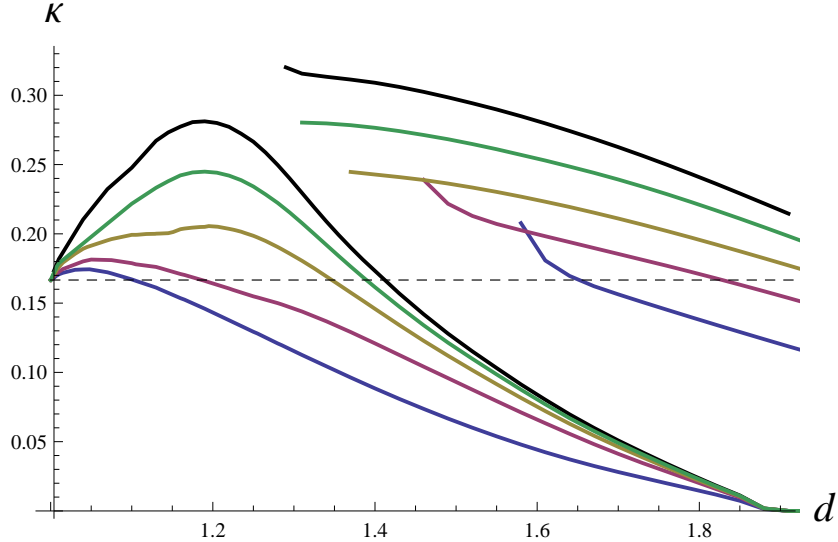}\\
\end{center}	
\caption{\label{fig:SONnogapGap} 
\small
Lower bounds on the two-point function coefficient $\kappa$ between two conserved SO$(N)$ or SU$(N/2)$ adjoint currents as obtained from a four-point function of scalar operators in the fundamental representation with dimension $d$ calculated at $k=10$. From below, the lines which start at $d=1$ correspond to $N=2$ (blue), $N=6$ (red), $N=10$ (brown), $N=14$ (green), $N=18$ (black), with no assumption on the spectrum. In the same order and using the same color code, the lines which start at $d\simeq 1.58 $, $d\simeq 1.46$, $d\simeq 1.37$, $d\simeq 1.31$ and $d\simeq 1.29$ show the bound which is obtained under the assumption that no scalar operator in the singlet channel has dimension $\Delta_S<4$. For illustrative purposes, we show the free-theory value $\kappa_{\rm free} = 1/6$ as a black dashed line.}
\end{figure}

In fig.~\ref{fig:SONnogapGap}, we report  our results in terms of lower bounds on $\kappa$. We have considered the five different values 
$N=2,6,10,14,18$  
and report the lower bounds on $\kappa$
for the case
where no assumption on the spectrum is made (the lines starting from $d=1$) and the case where the lowest-lying scalar operator in the singlet channel 
is assumed to have dimension $\Delta_S\geq 4$ (the other lines). The former bounds agree with previous results (e.g. compare with fig.~18 of ref.\cite{Poland:2011ey}). 
Although it is not clearly visible from the figure, we have checked that all the bounds consistently tend to the free-theory value for $d\rightarrow 1$.
The latter bounds start from a given $d_{\rm cr}> 1$ that depends on $N$. This is of course expected, given the known results for the upper bound on the dimension of the lowest-lying scalar singlet operator
at a given $d$: CFTs at $d<d_{\rm cr}$ are excluded under the assumption of a gap in the scalar singlet sector.
The values of $d_{\rm cr}$ that we find agree with the values given in the literature (compare e.g. with the dimensions $d$ for which $\Delta_0=4$ in fig.~4 of ref.\cite{Poland:2011ey}).
The lower bounds on $\kappa$ become significantly more stringent when we impose that $\Delta_S>4$. They also decrease less rapidly when $d$
increases compared to the unconstrained case.

\begin{figure}[!t]
\begin{center}
\hspace*{0cm} 
	\includegraphics[width=110mm]{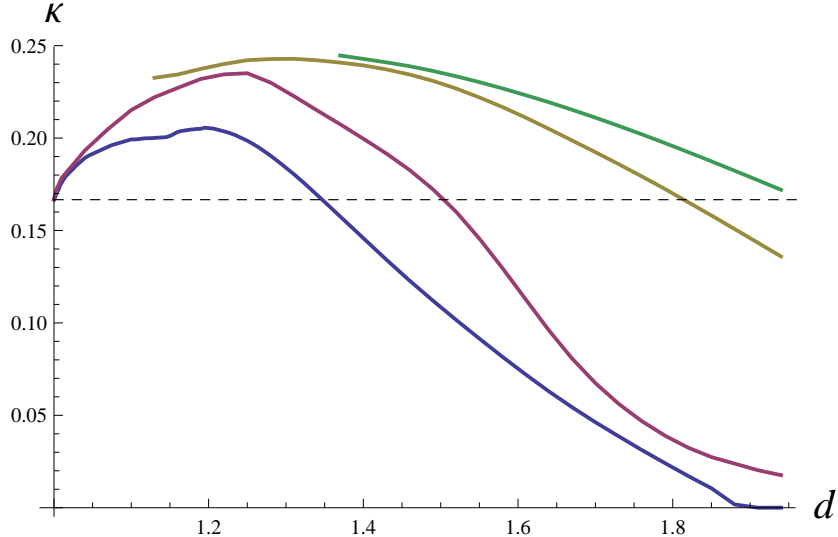}\\
\end{center}	
\caption{\label{fig:SO10NoGap_Gap123} 
\small
Lower bounds on the two-point function coefficient $\kappa$ between two conserved $SO(10)$ or $SU(5)$ adjoint currents as obtained from a four-point function of scalar operators in the fundamental representation with dimension $d$ calculated at $k=10$. From below, the lines correspond to the case with no assumption on the spectrum (blue) and assuming that no scalar operator in the singlet channel has dimension $\Delta_S<2$ (red),  $\Delta_S<3$ (brown), $\Delta_S<4$ (green). For illustrative purposes, we show the free-theory value $\kappa_{\rm free} = 1/6$ as a black dashed line.}
\end{figure}

In order to show how the assumption on $\Delta_S$ affects the lower bounds on $\kappa$, 
in fig.~\ref{fig:SO10NoGap_Gap123}, we fix $N=10$ and consider the three cases $\Delta_S\geq 2$,
$\Delta_S\geq 3$ and $\Delta_S\geq 4$. As expected, the lower bound consistently becomes more severe as we increase $\Delta_S$. As before, the bounds start at certain dimensions $d_{\rm cr}$ which agree with
previous results (compare e.g. with the dimensions $d$ for which $\Delta_0=3,4$ in fig.~4 of ref.\cite{Poland:2011ey}).

\subsection{SU$(N)$ Global Symmetry}
\label{sec:suN}

We consider a four-point function of complex scalars that are taken to be the components of a field in the fundamental representation of SU$(N)$ with dimension $d$.
The crossing symmetry relations have been derived in ref.\cite{Rattazzi:2010yc}. We report them here for completeness:
\be
\sum_{S^\pm} |\lambda_{\cal O}^S|^2 \!\left(\begin{array}{c}
F\\
H \\
(-)^l F \\
(-)^l H \\
0\\
0  \end{array}\right)
\!+\sum_{Ad^{\pm}} |\lambda_{\cal O}^{Ad}|^2\! \left(\begin{array}{c}
(1-\frac 1N) F \\
-(1+\frac 1N)H \\
(-)^{l+1}\frac 1N F \\ 
(-)^{l+1}\frac 1N H \\
(-1)^l F\\
(-)^l H \end{array}\right)
\!+\sum_{T^+} |\lambda_{\cal O}^T|^2\! \left(\begin{array}{c}
0 \\
0 \\
F\\ -H \\ F \\ -H\end{array}\right) 
\!+\sum_{A^-} |\lambda_{\cal O}^A|^2 \!\left(\begin{array}{c}
0 \\
0 \\
F\\ -H \\ -F \\ H\end{array}\right) 
\!=\! \left(\begin{array}{c}
1\\
-1 \\
1 \\
-1 \\
0\\
0 \end{array}\right).
\label{CrossingSUN}
\ee
Here $S$, $Ad$, $T$ and $A$ refer respectively to the singlet, adjoint, rank-2 symmetric and rank-2 antisymmetric representations of the operators ${\cal O}$ which define  the 
different conformal blocks. For the superscript $+$, even spins are included in the sum, and for $-$, odd spins are summed over.
We consider here the lower bounds on $\kappa$ (as defined in eq.~\eqref{kappadef}) associated with the adjoint current. 

As in subsection \ref{sec:soN}, we start by looking at the free theory of a complex scalar in the fundamental representation of SU$(N)$ in order to fix the constant $\rho$ in eq.~(\ref{lambdaKappa}).
The free-theory values of the OPE coefficients in the four different channels read
\be
\begin{split}
\lambda_{Ad,l}^{\rm free} & = \frac{1}{\sqrt{2}} \lambda_l^{\rm free}  \ \ \ \ \ \ (l\; {\rm even}\; and \; {\rm odd} )\,, \\ 
\lambda_{S,l}^{\rm free} & =\frac{1}{\sqrt{2N}}  \lambda_l^{\rm free}  \ \ \  (l\; {\rm even}\; and \; {\rm odd} )\ \,, \\
\lambda_{T,l}^{\rm free} & = \frac{1}{\sqrt{2}} \lambda_l^{\rm free} \ \ \ \ \ \ \ (l\; {\rm even}) \,, \\
\lambda_{A,l}^{\rm free} & =\frac{1}{\sqrt{2}} \lambda_l^{\rm free} \ \ \ \ \ \ \  (l\; {\rm odd}) \,,
\end{split}\ee
where $\lambda_l^{\rm free}$ is given in eq.~(\ref{Scalar1}). We in particular get $\lambda_J^{\rm free}=\lambda_{Ad,1}^{\rm free} = 1/\sqrt{2}$. 
Matching eq.~(\ref{betaCFT}) with the one-loop contribution to the $\beta$-function of a complex scalar in an SU$(N)$ gauge theory gives
\be
\kappa_{\rm free} =\frac 16 \,,
\label{KfreeSUN}
\ee
where we have taken $T({\rm fund.}) = 1/2$ (cf.~footnote \ref{kappafootnote}).
From this it follows that $\rho=1/12$  in eq.~(\ref{lambdaKappa}) as for SO$(N)$.

The six crossing symmetry equations (\ref{CrossingSUN}) should reduce to the three equations (\ref{CrossingSON}) when the group SU$(N)$ is embedded in an underlying SO$(2N)$ group.
The decomposition of the singlet, adjoint and rank-2 symmetric representations of SO$(2N)$ in terms of SU$(N)$ representations reads
\be\begin{split}
S_{{\rm SO}(2N)}^+ & = S_{{\rm SU}(N)}^+  \,, \\
T_{{\rm SO}(2N)}^+ & = T_{{\rm SU}(N)}^+\oplus \overline{T}^+_{{\rm SU}(N)}\oplus {\rm Ad}^+_{{\rm SU}(N)} \,, \\
A_{{\rm SO}(2N)}^-  & = A_{{\rm SU}(N)}^-\oplus \overline{A}^-_{{\rm SU}(N)}\oplus {\rm Ad}^-_{{\rm SU}(N)}\oplus S_{{\rm SU}(N)}^- \,.
\end{split} \label{deco}
\ee
If $\text{SU}(N)\subset\text{SO}(2N)$, for each primary operator in the $A^-$ ($T^+$) representation of SU$(N)$, there is a corresponding operator in the $Ad^-$ and $S^-$ ($Ad^+$) representation as follows from eq.~(\ref{deco}).
The OPE coefficients of these operators are related by the underlying SO$(2N)$ symmetry, $\lambda_{{\rm SO}(2N)}^{T^+} = \lambda_{{\rm SU}(N)}^{Ad^+}$, $\lambda_{{\rm SO}(2N)}^{A^-} = \lambda_{{\rm SU}(N)}^{A^-}=\sqrt{N}\lambda_{{\rm SU}(N)}^{S^-}$. It is straightforward to check with these identifications that eqs.~(\ref{CrossingSUN}) reduce to eqs.~(\ref{CrossingSON}).

As we have already mentioned, the numerical results for the lower bounds on $\kappa$ for SU$(N)$ are identical to those for SO$(2N)$, see fig.~\ref{fig:SONnogapGap}.
This suggests that, given a set of three functionals $\alpha_m$ that satisfy eq.~(\ref{alphaVect}) with $P=2$, $Q=1$ and $\eta_{F,H}$ as given by eq.~(\ref{CrossingSON}), one should be able to construct a set of six functionals $\tilde \alpha_m$ as linear combinations of the $\alpha_m$ such that these functionals satisfy eq.~(\ref{alphaVect}) with $P=3$, $Q=3$ and $\eta_{F,H}$ as given by eq.~(\ref{CrossingSUN}).
It would be interesting to find such a mapping and hence to understand in more analytical terms why the bounds on $\kappa$ for SO($2N)$ and SU($N$) are equal.

\subsection{$G_1\times G_2$ Global Symmetries}

The lower bounds on $\kappa$ found in subsections \ref{sec:soN} and \ref{sec:suN} apply to CFTs in presence of at least one scalar field in the fundamental representation of $G_1$, where $G_1=\text{SO}(M)$ or SU$(M)$.
Of course, the CFT can contain additional charged fields, for example a number $N$ of scalars in the fundamental representation of $G_1$, with dimensions $d_1,\ldots d_N$. 
In the free-theory limit of $N$ decoupled scalars (real for SO$(M)$, complex for SU$(M)$) we would simply have
\be
\kappa_{\rm free}=\frac N6\,.
\ee
The larger $N$ is, the more  constraining (and interesting) the lower bounds are.
One cannot naively rescale the results of fig.~\ref{fig:SONnogapGap} by a factor of $N$ in order to match the new free theory limit, however, because
the interactions among the scalars will not be taken into account in this way.
A more constraining bound could likely be obtained by studying the coupled set of four-point functions involving all $N$ scalars.
This is in general not straightforward to do, since the crossing symmetry constraints are significantly more involved in presence of fields with different scaling dimensions.\footnote{See ref.\cite{Kos:2014bka} for very recent progress in this direction.}
A simple way to mimic the presence of more fields charged under a given group, though at the cost of assuming identical scaling dimensions $d_1=\ldots d_N=d$, is obtained by introducing a further global symmetry group
$G_2$  and assuming that the $N$ fields transform under some representation of $G_2$.
This is the main motivation for us to consider global symmetries which are direct products of two simple groups: it is a way to obtain lower bounds on $\kappa_{G_1}$ in presence of more than one field
charged under $G_1$. More specifically, in the following we will consider fields in the fundamental representation of $G_2=\text{SO}(N)$.

 \begin{figure}[!t]
\begin{center}
\hspace*{0cm} 
\includegraphics[width=120mm]{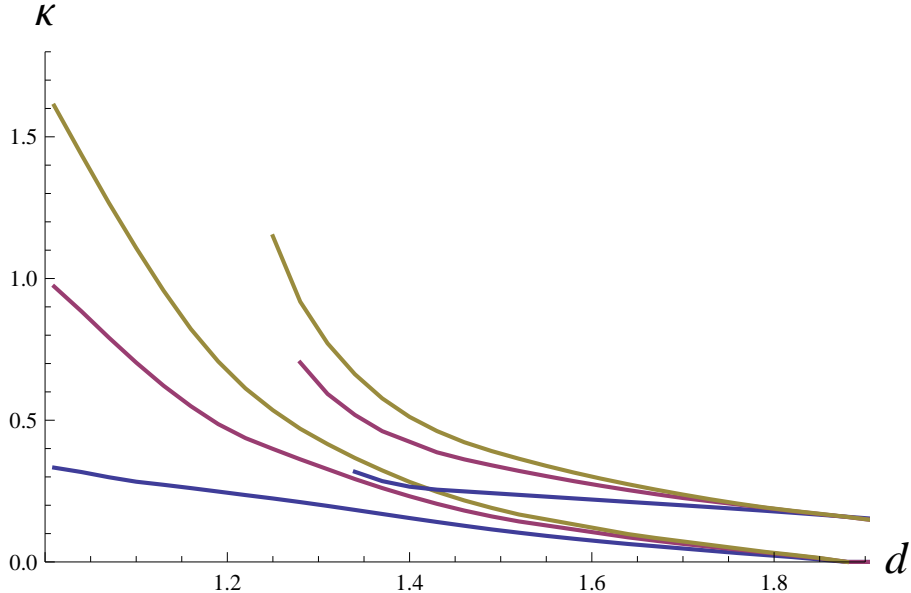}\\
\end{center}
\caption{\label{fig:SONSOMnogapGap} 
\small
Lower bounds on the two-point function coefficient $\kappa$ between two conserved SO$(N)$ (or SU$(N/2)$) adjoint currents as obtained from a four-point function of scalar operators with dimension $d$ in the 
bi-fundamental representation of SO$(N)\times$SO$(M)$,  calculated at $k=9$. We take $N=6$.
From below, the lines which start at $d=1$ correspond to $M=2$ (blue), $M=6$ (red), $M=10$ (brown), with no assumption on the spectrum. In the same order
and using the same color code, the lines which start at $d\simeq 1.34 $, $d\simeq 1.28$ and $d\simeq 1.25$ show the bound which is obtained under the assumption that no scalar operator in the singlet channel has dimension $\Delta_S<4$.}
\end{figure}

\subsubsection{SO$(N)\times$SO$(M)$}

Consider a CFT with global symmetry SO$(N)\times$SO$(M)$ and one real scalar $\phi_a^i$ in the bi-fundamental representation of SO$(N)\times$SO$(M)$, 
$a$ and $i$ being SO$(N)$ and SO$(M)$ indices, respectively.
In complete analogy to the SO$(M)$ case discussed in ref.\cite{Rattazzi:2010yc},  we can  impose crossing symmetry in the $s$- and $t$-channel on the 
four-point function $\langle \phi_a^i(x_1)  \phi_b^j(x_2)  \phi_c^k(x_3)  \phi_d^l(x_4) \rangle$ in order to obtain the bootstrap equations.
The operators appearing in the $\phi\phi$ OPE transform under SO$(N)\times$SO$(M)$ according to the decomposition of $({\bf N},{\bf M})\otimes ({\bf N},{\bf M})$, where {\bf N} and {\bf M} denote the fundamental representations of respectively SO$(N)$ and SO$(M)$.
This gives 9 different representations, consisting of pairs $(ij)$, where $i,j=S,T,A$ refer to the singlet ($S$), symmetric ($T$) and antisymmetric ($A$) representations
of respectively SO$(N)$ and SO$(M)$. Correspondingly, we get a total of $3\times 3 = 9$ equations. We report them in eq.~(\ref{sonsom}) in appendix \ref{subsec:sonsom}.

\begin{figure}[!t]
\begin{center}
\hspace*{0cm} 
\includegraphics[width=120mm]{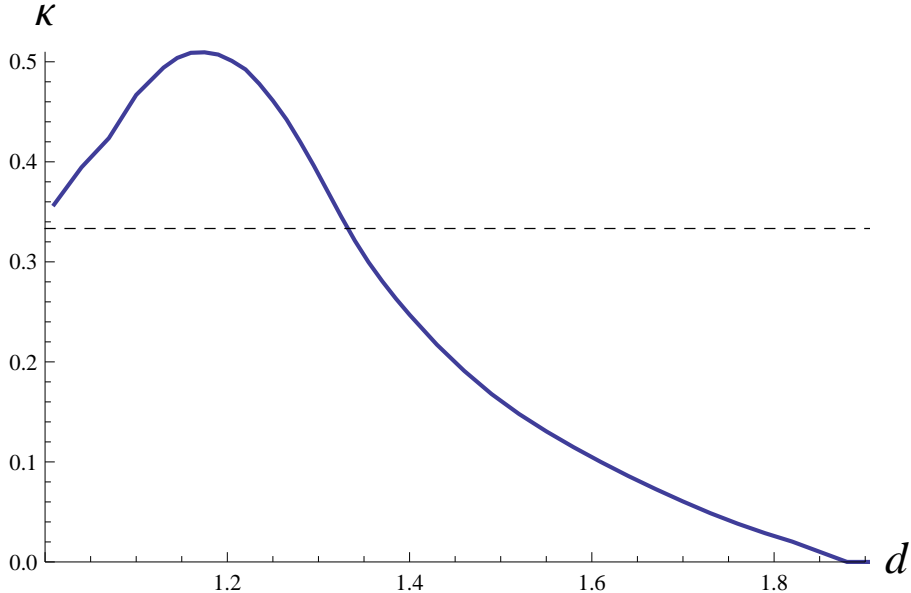}\\
\end{center}
\caption{\label{fig:SO30SO2} 
\small
Lower bounds on the two-point function coefficient $\kappa$ between two conserved SO$(30)$ (or SU$(15)$) adjoint currents as obtained from a four-point function of scalar operators with dimension $d$ in the 
bi-fundamental representation of SO$(30)\times$SO$(2)$, calculated at $k=9$. No assumption on the spectrum is made. The black dashed line corresponds to the free-theory value $\kappa_{\rm free} = 1/3$.}
\end{figure}

The SO$(N)$ conserved current that we analyze is in the $(AS)$ representation and is the lowest-dimensional operator appearing in the functions $F_{AS}$ and $H_{AS}$ defined in eq.~(\ref{FHijso}). In fig.~\ref{fig:SONSOMnogapGap}, we show the lower bounds on $\kappa_{{\rm SO}(6)}$ for the three cases $M=2,6,10$. 
While for the case of SO$(N)$ or SU$(N/2)$ considered before the lower bound
first becomes significantly more stringent with growing $d$ and only from a certain $d$ onwards becomes less stringent, here a slight increase in the bound arises only for $d$ very close to 1 after which the bound decreases with $d$.
The lines starting from $d=1$ correspond to the case where no assumption on the spectrum has been made,
while for the other lines we have assumed that the lowest-lying scalar operator in the SS channel 
has dimension $\Delta_{SS}\geq 4$. The lower bounds for the latter case are stronger than for the former, but the difference is less substantial than for the groups SO$(N)$ or SU$(N/2)$.
The lower bound $d\geq d_{\rm cr}$ on the dimension of $\phi_a^i$ above which the lowest-lying operator in the $SS$ channel can have a dimension $\Delta_{SS}\geq 4$
is also weaker than what was found for SO$(N)$ or SU$(N/2)$.
This is expected, since this bound becomes the weaker the larger the group is.
The correct free-theory limit is obtained in all three cases.
The shape of the lower bound on $\kappa$ with no assumption on the spectrum in fig.~\ref{fig:SONSOMnogapGap} resembles the bound found in ref.\cite{Poland:2011ey} for SU$(N)$ singlet currents (see e.g. their fig.~19). From the SO$(M)$ point of view, the SO$(N)$ current is in fact a collection of $N(N-1)/2$ singlet currents. On the other hand, for $N\gg M$,  we find that the lower bound on $\kappa$ for SO$(N)$ currents shows the characteristic bump of single SO$(N)$ or SU$(N/2)$ currents, well above the free-theory value, as in fig.~\ref{fig:SONnogapGap} (a). 
For illustration, we show the bound on $\kappa$ obtained for $N=30$ and $M=2$ in fig.~\ref{fig:SO30SO2}. 
It would be interesting to further explore these bounds and to understand the origin of their different behaviours in the regimes $N\leq M$ and $N\gg M$.

\begin{figure}[!t]
\begin{center}
\hspace*{0cm} 
\includegraphics[width=120mm]{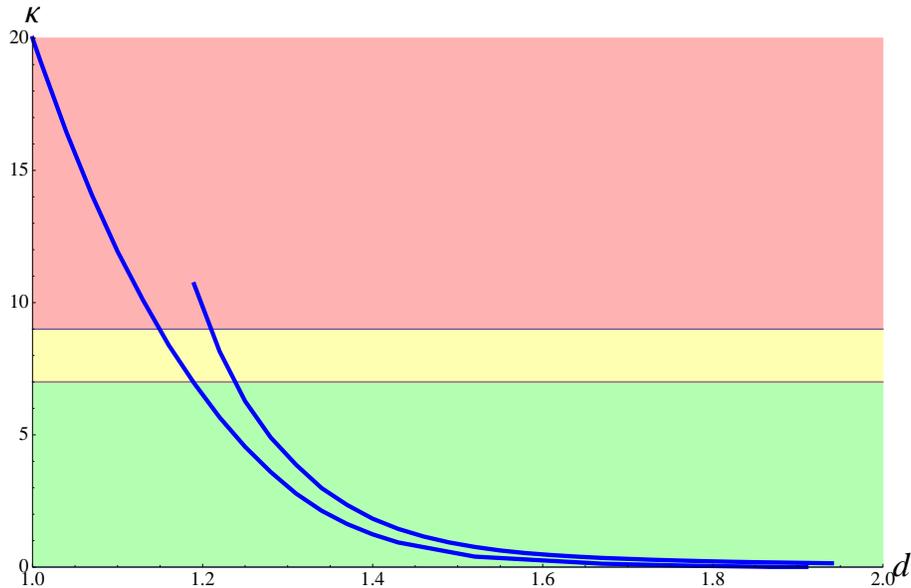}\\
\end{center}
\caption{\label{fig:SO120} 
\small
Lower bounds on the two-point function coefficient $\kappa$ between two conserved $\text{SO}(6)\supset\text{SU}(3)_c$  adjoint currents as obtained from a four-point function of scalar operators with dimension $d$  in the 
bi-fundamental representation of SO$(6)\times$SO$(120)$, calculated at $k=9$. 
The line which starts at $d=1$ corresponds to the case where no assumption is made on the spectrum, whereas for the line which starts at $d\simeq 1.19$, the CFT is assumed to have no scalar operator in the singlet channel with dimension $\Delta_S<4$.
In the green region $\alpha_c$ remains asymptotically free, while in the orange and red regions $\alpha_c$ develops trans-Planckian and sub-Planckian Landau poles, respectively.}
\end{figure}

As a final application of our results, in fig.~\ref{fig:SO120}, we report the lower bounds on $\kappa$ for the group SO$(6)\times$SO$(120)$. We choose SO$(120)$ because 
the contribution of 120 free complex scalar triplets to the $\text{SU}(3)_c\subset\text{SO}(6)$ current-current two-point function gives $\kappa= 20$.
This in turn is the same value found in eq.~(\ref{kSO5free}) for the number of free fermion triplets which are needed to give mass to all the SM quarks in the SO(5)$\rightarrow$SO$(4)$ pNGB composite Higgs model mentioned in section \ref{sec:Pheno}.  We consider SO(6)$\times$SO$(120)$ and not SU$(3)\times$SO$(120)$ because the latter case is computationally very demanding (incidentally, 
in one of the models presented in ref.\cite{Caracciolo:2012je}, SU$(3)_c$ was actually embedded in an underlying SO$(6)$ flavour global symmetry). Anyhow, given the equivalence
between the SO$(2N)$ and SU$(N)$ lower bounds on $\kappa$, we believe that these results would also hold for the SU(3)$\times$SO$(120)$ case.

As we see in fig.~\ref{fig:SO120}, assuming the absence of a relevant scalar singlet operator in the CFT
does not significantly change the bounds.
Furthermore, the most dangerous region regarding Landau poles, which is the region close to $d=1$, is not consistent with the assumption of absence of relevant deformations. If we demand that no sub-Planckian Landau pole arises, then we need $d\gtrsim 1.2$, while
for $d\gtrsim 1.25$, $\alpha_c$ remains asymptotically free.

\subsubsection{SO$(N)\times$SU$(M)$}

Consider a CFT with global symmetry SO$(N)\times$SU$(M)$ and one complex scalar $\phi_a^i$ in the bi-fundamental representation of SO$(N)\times$SU$(M)$, 
$a$ and $i$ being SO$(N)$ and SU$(M)$ indices, respectively. We impose crossing symmetry in the $s$- and $t$-channel on the 
four-point function $\langle \phi_a^i(x_1)  \phi_b^{\bar j\dagger}(x_2)  \phi_c^k(x_3)  \phi_d^{\bar l\dagger}(x_4) \rangle$ and the four-point function with $x_3\leftrightarrow x_4$.
The operators appearing in the $\phi\phi$ OPE transform under SO$(N)\times$SU$(M)$ in representations $(ij)$, where $i=S,T,A$ refer to the singlet ($S$), symmetric ($T$) and antisymmetric ($A$) representations
of SO$(N)$ and $j=A,T$ refer to the symmetric ($T$) and antisymmetric ($A$) representations of SU$(M)$. This gives 6 different representations, 
with even and/or odd spin operators, depending on the representation.
The operators appearing in the $\phi\phi^\dagger$ OPE transform in representations $(ij)$, with $i=S,T,A$ as before, 
whereas $j=S,Ad$ refer to the singlet ($S$) and adjoint ($Ad$) representations of SU$(M)$. Taking into account that SU$(M)$ singlet and adjoint operators appear with both even and odd spins, we get 12 
different conformal blocks, for a total of 18 bootstrap equations. We report them in eqs.~(\ref{sonsump1}) -- (\ref{sonsump2b}) in appendix \ref{subsec:sunsom}.
The SO$(N)$ conserved current that we are interested in transforms under the $AS$ representation and is the lowest-dimensional operator appearing in the functions $F_{AS}^-$ and $H_{AS}^-$ defined in eq.~(\ref{FHijsu}).

The large number of bootstrap equations makes the numerical analysis of the SO$(N)\times$SU$(M)$ case computationally very demanding.
We have however been able to compute the lower bounds on $\kappa_{SU(M)}$ at level $k=6$ with $L=15$, using $N=30$ and $M=3$
and either assuming that no scalar operator has dimension $\Delta_S<4$ or without any assumption.
The results are essentially indistinguishable from those obtained for $\kappa_{SO(6)}$ when considering SO$(N)\times$SO$(2M)$, though
the accuracy obtained is not enough to claim that they are identical. Given also the observed equivalence of the bounds for SU$(M)$ and SO$(2M)$, we take  this result as enough evidence to conjecture that the bounds for SO$(N)\times$SU$(M)$ are identical to those for SO$(N)\times$SO$(2M)$, for any $N$ and $M$.

\section{Conclusions}

In this paper, we have numerically studied bounds on various OPE coefficients in 4D CFTs using the bootstrap approach applied to four-point functions of scalar operators.  We have first studied bounds on OPE coefficients of symmetric traceless tensor operators with spins $l=2$ and $l=4$ as a function of their scaling dimension.  Furthermore, we have analyzed how an assumption on the dimension of the lowest-lying scalar operator affects such bounds.

We have then considered 4D CFTs with a global symmetry $G$. 
When this group, or a subgroup of it, is gauged by weakly coupling external gauge fields to the CFT, the coefficient $\kappa$ which enters in the two-point function of the associated conserved vector currents
governs the leading CFT contribution to the one-loop $\beta$-function of the corresponding gauge coupling. In particular, if this contribution is too large, it gives rise to unwanted sub-Planckian Landau poles. Motivated by physics beyond the Standard Model, where $G_{\rm SM} \subseteq G$, we have numerically studied 
the lower bounds on the coefficient $\kappa$ using techniques developed in ref.\cite{Poland:2011ey}. 
Possible hierarchy problems are avoided by demanding that all scalar operators in the spectrum which are singlets under the global symmetry
have dimensions $\Delta_S\geq 4$. More specifically, we have considered lower bounds on $\kappa$ extracted from crossing symmetry constraints on four-point functions
of scalar operators $\phi_i$ in the fundamental representation of SO$(N)$, or the bi-fundamental representation of SO$(N)\times$SO$(M)$.

We have been mostly motivated by applications in the context of composite Higgs models with partial compositeness, where the CFT is assumed to have a global symmetry $G$
and a set of fermion operators with different dimensions. Our results are encouraging. For concreteness, we have considered a CFT with $\text{SU}(3)_c\subset\text{SO}(6)$ global symmetry. We have chosen the scalar matter content  of the CFT such that, in the free-theory limit, it has the same quantitative effect on the $\beta$-function of $\alpha_c$ as the fermion matter content in a popular composite pNGB Higgs model based on the coset SO(5)/SO(4).
In this setting, Landau poles for $\alpha_c$ can always be avoided as long as the dimension of the external scalar operator is not too close to the limiting value set by demanding that $\Delta_S\geq4$ (see fig.~\ref{fig:SO120}). Of course, the bootstrap constraints that we have obtained are only lower bounds, and it is well possible that these bounds can be significantly improved. For example, 
imposing the bootstrap equations for several field multiplets directly without invoking an additional symmetry, along the lines of ref.\cite{Kos:2014bka},  might lead to stronger bounds compared to the rather weak bounds on SO(N) currents that we have found for CFTs with SO(N)$\times$SO(M) global symmetry when $N\leq M$.
Our results are obviously not directly applicable to the actual phenomenological models, because i) we have considered four-point functions of scalars and not fermions
and ii) we have assumed that the scalars all have the same scaling dimension.  Nevertheless, we believe that these preliminary results can be a useful first step towards 
a more comprehensive analysis that goes beyond these two limitations.

\section*{Acknowledgments}

We thank Alfredo Urbano, Balt van Rees, David Poland and especially Slava Rychkov for useful discussions and correspondence.  
We are particularly grateful to the INFN technical staff of the Zefiro cluster in Pisa, where most of the computations for this paper have been performed. 
B.v.H. and M.S. thank the Galileo Galilei Institute for Theoretical
Physics for hospitality during the completion of this work.

\appendix

\section{Details about the Numerical Procedure}
\label{app:sdpa}

In order to discuss the numerical procedure developed in ref.\cite{Poland:2011ey}, let us consider the simplest case of an external singlet operator. The constraints that the functional $\alpha$ needs to fulfill in order to get bounds on OPE coefficients are given in eq.~\eqref{alpha}. It is convenient to first rescale the bootstrap equation by a $(\Delta,l)$-independent function $f(z,\bar{z})$ (see ref.\cite{Poland:2011ey} for more details). 
In particular, the positivity constraints on the rescaled conformal blocks $E^+_{d,\Delta,l}\equiv f(z,\bar{z}) F_{d,\Delta,l}$ then read
\be
a_{m n} \, \partial^m_z \partial_{\bar{z}}^n \, E^+_{d,\Delta,l} \, \geq \, 0\,  \quad \quad  \forall (\Delta,l)\neq (\Delta_0,l_0) \, ,
\label{posconstr}
\ee
where summation over $m$ and $n$ is understood and the derivatives are evaluated at $z=\bar{z}=1/2$. The crucial insight is that the derivatives of $E^+_{d,\Delta,l}$ allow for an approximation
\be
 \partial^m_z \partial_{\bar{z}}^n \, E^+_{d,\Delta,l} \, \simeq \, \chi_{l}(\Delta) U^{mn}_{l,d,+}(\Delta)\,,
\label{conblockapprox}
\ee
where $\chi_{l}(\Delta)$ is a positive definite function of $\Delta$ and $U^{mn}_{l,d,+}(\Delta)$ is a polynomial in $\Delta$. We use 5 roots for this approximation (see ref.\cite{Poland:2011ey} for more details). An analogous approximation can be found for the rescaled conformal blocks $U^{mn}_{l,d,-}(\Delta) \equiv \tilde{f}(z,\bar{z}) H_{d,\Delta,l}$ that appear when dealing with global symmetries. Making use of a theorem by Hilbert, the positivity constraints in eq.~\eqref{posconstr} can equivalently be formulated as the requirement that there exist positive semidefinite matrices $A_l$ and $B_l$ such that
\be
a_{mn} \, U^{mn}_{l,d,+}(\Delta_l (1+x)) \, = \,  X_p \, A_l \, X_p^T \, + \, x \, X_q \, B_l \, X_q^T \quad \quad  \forall l\neq l_0 \, .
\label{posconstrrew}
\ee
Here $X_p \equiv (1,x,...,x^p)$ is a vector and $p$ and $q$ are determined by the degree of the polynomial $U^{mn}_{l,d,+}$. Furthermore, $\Delta_l$ is the unitarity bound on the operator dimension. The task now consists of finding coefficients $a_{mn}$ and a set of matrices $A_l$ and $B_l$ such that eq.~\eqref{posconstrrew} is fulfilled  (additional constraints arise from e.g.~the normalization condition $\alpha(F_{d,\Delta_0,l_0}) = 1$ in eq.~\eqref{alpha} and the minimization of $\alpha(1)$ in eq.~\eqref{boundequation}). This can be formulated as a positive semidefinite program for which there exist powerful numerical codes. 
The existence of such coefficients and matrices guarantees the positivity of the functional for all $\Delta \geq \Delta_l$ (corresponding to $x \geq 0$). 
As already discussed in section \ref{sec:Num}, on the other hand, we can only take a finite number of spins $l$ into account. 

We use Mathematica 9.0 to calculate the coefficients of the polynomials $U^{mn}_{l,d,\pm}$ and to set up the positive semidefinite program. The data is written to file and handed to the numerical code SDPA-GMP 7.1.2 \cite{sdpa} (using their sparse data format) which solves the positive semidefinite program. We use the same parameter set for the SDPA-GMP as ref.\cite{Poland:2011ey} (see the table in their Appendix B). For the calculations, we have used the Zefiro cluster of the INFN which is located in Pisa (Italy). This cluster consists of 25 computers, each of which has 512 GB RAM and 4 processors with 16 cores. For the plots, we have calculated points with a spacing of $\delta d=3\cdot10^{-2}$ or $\delta\Delta=3\cdot10^{-2}$. In order to obtain smooth plots, we interpolate between these points.

The Multiple Precision Arithmetic Library (GMP) allows to carry out calculations up to high precision.
This is necessary because the numerical values of the coefficients of the polynomials $U^{mn}_{l,d,\pm}$ span several orders of magnitudes.
An important source for this spread are conformal blocks with large spins $l\gtrsim 10$. For these values of $l$, an asymptotic expression for the conformal blocks and its derivatives is 
a good approximation. Taking $\smash{z=1/2 + a+b}$ and $\smash{\bar z = 1/2 +a -b}$,
for $l^2\gg \Delta-l-2$ one finds \cite{Rattazzi:2008pe}
\be
\partial_a^{2m}\partial_b^{2n} F_{d,\Delta,l}|_{a=b=0} \, \simeq \, \frac{{\rm const.}}{(2m+1)(2n+1)} (2\sqrt{2}l)^{2m+2n+2} e^{-c \,  l}\,,
\label{FAsym}
\ee
where $c=-\log(12-8\sqrt{2})\simeq 0.377$ and const.~is a positive constant of ${\cal O}(1)$ that only depends on $d$.
A straightforward generalization of this result allows to also find an asymptotic analytic expression for the conformal block $H$ defined in eq.~(\ref{HDef}):
\be
\partial_a^{2m}\partial_b^{2n} H_{d,\Delta,l}|_{a=b=0}\, \simeq \, \frac{{\rm const.}}{(2n+1)} (2\sqrt{2}l)^{2m+2n+1} e^{-c \, l}\,.
\label{HAsym}
\ee
From the above two results, we find that the spread among the coefficients 
of the polynomials for a given spin $l \gtrsim 10$ is at least of order ${\cal O}(l^{2k+2})$ for conformal blocks $F$ and ${\cal O}(l^{2k+1})$ for $H$. 
In addition, these results allow us to estimate the value $l_{\rm max}$ for which derivatives of the conformal blocks have a maximum (in which case potential violations of the positivity constraint in eq.~\eqref{alpha} could give a large correction in eq.~\eqref{boundequation}). To this end, notice that for a given $l$, the largest coefficients arise from the highest derivatives with $m+n<2k$. Maximizing these coefficients with respect to $l$ then yields the formula for $l_{\rm max}$ in eq.~(\ref{lD}).

An additional source for the spread arises from the approximation in eq.~\eqref{conblockapprox}. The functions $\chi_l(\Delta)$ are numerically small for large spins $l$ and therefore increase the spread among the various coefficients of the polynomials  $U^{mn}_{l,d,\pm}(\Delta) \, \simeq \, \partial^m_z \partial_{\bar{z}}^n \, E^\pm_{d,\Delta,l} \, / \, \chi_{l}(\Delta)$ that determine the  positive semidefinite program. 

In order to reduce the numerical spread among the polynomial coefficients (which allows to reduce the required precision and thereby speeds up the calculation), we rescale them by both an $(m,n)$-dependent factor and an $l$-dependent factor before handing them to the SDPA-GMP. Both of these rescalings transform the positivity constraint eq.~\eqref{posconstrrew} into an equivalent constraint. Indeed, the $(m,n)$-dependent factor amounts to a redefinition of the coefficients $a_{mn}$, whereas the $l$-dependent rescaling can be absorbed into the matrices $A_l$ and $B_l$. Note, however, that e.g.~the effect of the rescaling on the normalization condition $\alpha(F_{d,\Delta_0,l_0}) = 1$ in eq.~\eqref{alpha} needs to be taken into account when calculating the bound from eq.~\eqref{boundequation}.


\section{Crossing Relations for SO$(N)\times$SO$(M)$ and SO$(N)\times$SU$(M)$}
\label{app:crossing}

We report here the crossing symmetry constraints coming from four-point functions of scalar operators with scaling dimensions $d$ in the bi-fundamental representation of 
SO$(N)\times$SO$(M)$ and SU$(N)\times$SO$(M)$.

\subsection{SO$(N)\times$SO$(M)$}
\label{subsec:sonsom}

Let $\phi_a^i$ be the scalar operator in the bi-fundamental representation of SO$(N)\times$SO$(M)$, with $a$ and $i$ being SO$(N)$ and SO$(M)$ indices, respectively.
As usual, we define conformal blocks that contain the contributions of the operators appearing in the OPE of $\phi_a^i \phi_b^j$  in a given representation of the global symmetry.
We have nine different conformal blocks $G_{ij}$, where $i,j=S,T,A$ with $S$, $T$ and $A$ corresponding to the singlet, symmetric and antisymmetric representations
of SO$(N)$ and SO$(M)$. The first index refers to SO$(N)$, the second one to SO$(M)$. The spin of the operators entering in $G_{ij}$ is even if zero or two antisymmetric representations appear and
odd otherwise. In order to have reasonably compact formulas, we define the functions
\be
F_{ij}  \equiv \hspace{-.4cm} \sum_{{\cal O}\in (i,j){\rm -sector}} \hspace{-.4cm} |\lambda_{{\cal O}}^{ij}|^2 \, \, F_{d,\Delta,l}(z,\bar z) \,, \ \ \ \ H_{ij}  \equiv \hspace{-.4cm} \sum_{{\cal O}\in (i,j){\rm -sector}} \hspace{-.4cm}  |\lambda_{{\cal O}}^{ij}|^2 \, \,  H_{d,\Delta,l}(z,\bar z) \,.
\label{FHijso}
\ee
In terms of these, the crossing relations read
\be\begin{split}
& F_{SS}-\frac 2M F_{ST}-\frac 2N F_{TS} +\Big(1+\frac{4}{MN}\Big)F_{TT}+F_{AT}+F_{TA}+F_{AA} = 1\,, \\
& H_{SS}-\frac 2M H_{ST}-\frac 2N H_{TS} -\Big(1-\frac{4}{MN}\Big)H_{TT} -H_{AT}-H_{TA}-H_{AA} = -1\,,  \\
&  \Big(1-\frac{2}{M}\Big) F_{TT}  +F_{TS}-F_{AS}+F_{TA}-\Big(1-\frac{2}{M}\Big)F_{AT} -F_{AA}=0\,, \\
&  \Big(1+\frac{2}{M}\Big) H_{TT}  -H_{TS}+H_{AS}+H_{TA}-\Big(1+\frac{2}{M}\Big)H_{AT} -H_{AA}=0\,, \\
&  \Big(1-\frac{2}{N}\Big) F_{TT}  +F_{ST}-F_{SA}+F_{AT}-\Big(1-\frac{2}{N}\Big)F_{TA} -F_{AA}=0\,, \\
&  \Big(1+\frac{2}{N}\Big) H_{TT}  -H_{ST}+H_{SA}+H_{AT}-\Big(1+\frac{2}{N}\Big)H_{TA} -H_{AA}=0\,, \\
&  \Big(\frac{2}{M}+\frac 2N\Big) F_{TT} + \frac 2N F_{TA}+\frac 2M F_{AT}-F_{TS}-F_{ST} -F_{SA}-F_{AS}=0\,, \\
&  \Big(\frac{2}{M}-\frac 2N\Big) H_{TT} - \frac 2N H_{TA}+\frac 2M H_{AT}-H_{TS}+H_{ST} +H_{SA}-H_{AS}=0\,,\\
& F_{TT}-F_{AT}-F_{TA}+F_{AA} = 0\,.
\label{sonsom}
\end{split}\ee
We have verified that reflection positivity is satisfied in the appropriate channels. 
The values of the OPE coefficients in the free-theory limit $d\rightarrow 1$ read
\be\begin{split}
&\lambda^{TT}_l = \lambda^{AA}_l= \lambda^{AT}_l= \lambda^{TA}_l = \frac 12 \lambda_l^{free}\,,  \\
&\lambda^{TS}_l = \lambda^{AS}_l= \frac{1}{\sqrt{2M}}  \lambda_l^{free}\,,  \\
&\lambda^{ST}_l = \lambda^{SA}_l= \frac{1}{\sqrt{2N}}  \lambda_l^{free}\,,  \\
&\lambda^{SS}_l= \frac 1{2\sqrt{MN}} \lambda_l^{free} \,,
\label{lambdasonsom}
\end{split}\ee
where $\lambda_l^{\rm free}$ is given in eq.~(\ref{Scalar1}) and $l$ is even or odd depending on the representation. Consistency with the free-theory limit provides a further check on various signs appearing in eq.~(\ref{sonsom}).

\subsection{SO$(N)\times$SU$(M)$}
\label{subsec:sunsom}

Let $\phi_a^i$ and $\phi_a^{\bar i,\dagger}$ be a scalar operator and its complex conjugate in the bi-fundamental representation of SO$(N)\times$SU$(M)$, with $a$ and $i$ being SO$(N)$ and SU$(M)$ indices, respectively. As usual, we define conformal blocks that contain the contributions of the operators appearing in the OPE of $\phi_a^i \phi_b^j$  in a given representation of the global symmetry.
Since operators in the singlet and adjoint representations of  SU$(M)$ can have both even and odd spin, we define
\be\begin{split}
& F_{ij}^{+/-}  \equiv \hspace{-.4cm} \sum_{\substack{{\cal O}\in (i,j) {\rm -sector} \\ l \, \,  {\rm even/odd}}} \hspace{-.4cm} |\lambda_{{\cal O}}^{ij_{+/-}}|^2 \, \, F_{d,\Delta,l}(z,\bar z) \,, \ \ \ \ \ \ H_{ij}^{+/-}  \equiv \hspace{-.4cm} \sum_{\substack{{\cal O}\in (i,j) {\rm -sector}\\ l \, \, {\rm even/odd}}} \hspace{-.4cm}  |\lambda_{{\cal O}}^{ij_{+/-}}|^2 \, \, H_{d,\Delta,l}(z,\bar z) \,, \\
& F_{ij} \equiv F_{ij}^{+}+F_{ij}^{-}\,, \quad \hat F_{ij} \equiv F_{ij}^{+}-F_{ij}^{-} \,, \quad \quad  \  \, H_{ij} \equiv H_{ij}^{+}+H_{ij}^{-}\,, \quad \hat H_{ij} \equiv H_{ij}^{+}-H_{ij}^{-} \,.
\label{FHijsu}
\end{split}\ee
Here $i$ runs over the representations $S,T,A$ of SO$(N)$, while $j$ runs over the singlet $(S)$, adjoint $(Ad)$, symmetric $(T)$ and antisymmetric $(A)$ representations of SU$(M)$.
Distinguishing between even and odd spins, we have a total of 18 conformal blocks and, correspondingly, a system of 18 crossing symmetry constraints. Six of these constraints arise by imposing
crossing symmetry in the $s$- and $t$-channel on the four-point function $\langle \phi_a^i \phi_b^{\bar j,\dagger} \phi_c^k \phi_d^{\bar l,\dagger}\rangle$. They read
\be\begin{split}
& F_{SS}-\frac 2N F_{TS}-\frac 1M F_{SAd}+\Big(1+\frac{2}{MN}\Big) F_{TAd}+F_{AAd} = 1\,, \\
& H_{SS}-\frac 2N H_{TS}-\frac 1M H_{SAd}-\Big(1-\frac{2}{MN}\Big) H_{TAd}-H_{AAd} = -1\,, \\
& F_{TS}-F_{AS}+\Big(1-\frac 1M\Big) F_{TAd}-\Big(1-\frac 1M\Big) F_{AAd} = 0\,, \\
& H_{TS}-H_{AS}-\Big(1+\frac 1M\Big) H_{TAd}+\Big(1+\frac 1M\Big) H_{AAd} = 0\,, \\
& F_{TS}+F_{AS}-\Big(\frac 1M+\frac 2N\Big) F_{TAd}+F_{SAd}-\frac 1M F_{AAd} = 0\,, \\
& H_{TS}+H_{AS}+\Big(-\frac 1M+\frac 2N\Big) H_{TAd}-H_{SAd}-\frac 1M H_{AAd} = 0\,.
\label{sonsump1}
\end{split}\ee
The remaining twelve constraints arise by imposing crossing symmetry in the $s$- and $t$-channel on the four-point function $\langle \phi_a^i \phi_b^{\bar j,\dagger} \phi_c^{\bar k,\dagger} \phi_d^l\rangle$.
They read
\be\begin{split}
& \hat F_{SS}-\frac 2N \hat F_{TS}-\frac 1M \hat F_{SAd}+\frac{2}{MN} \hat F_{TAd}+F_{TT}^++F_{AT}^-+F_{TA}^-+F_{AA}^+= 1\,, \\
& \hat H_{SS}-\frac 2N \hat H_{TS}-\frac 1M \hat H_{SAd}+\frac{2}{MN} \hat H_{TAd}-H_{TT}^+-H_{AT}^--H_{TA}^--H_{AA}^+= -1\,, \\
& \hat F_{TS}-\frac 1M \hat F_{TAd}-\frac{1}{M} \hat F_{AAd}+\hat F_{AS}+F_{TT}^+-F_{AT}^-+F_{TA}^--F_{AA}^+= 0\,, \\
& \hat H_{TS}-\frac 1M \hat H_{TAd}-\frac{1}{M} \hat H_{AAd}+\hat H_{AS}-H_{TT}^++H_{AT}^--H_{TA}^-+H_{AA}^+= 0\,, \\
& \hat F_{TS}-\frac 1M \hat F_{TAd}+\frac{1}{M} \hat F_{AAd}-\hat F_{AS}-\frac 2N F_{TT}^++F_{ST}^++F_{SA}^--\frac 2N F_{TA}^-= 0\,, \\
& \hat H_{TS}-\frac 1M \hat H_{TAd}+\frac{1}{M} \hat H_{AAd}-\hat H_{AS}+\frac 2N H_{TT}^+-H_{ST}^+-H_{SA}^-+\frac 2N H_{TA}^-= 0\,, \\
& \hat F_{SAd}-\frac 2N  \hat F_{TAd}+ F_{TT}^+-F_{TA}^-+F_{AT}^--F_{AA}^+= 0\,, \\
& \hat H_{SAd}-\frac 2N  \hat H_{TAd}-H_{TT}^++H_{TA}^--H_{AT}^-+H_{AA}^+= 0\,, \\
& \hat F_{TAd}+ \hat F_{AAd}+ F_{TT}^+-F_{TA}^--F_{AT}^-+F_{AA}^+= 0\,, 
\label{sonsump2a}
\end{split}\ee
\be\begin{split}
& \hat H_{TAd}+ \hat H_{AAd}-H_{TT}^++H_{TA}^-+H_{AT}^--H_{AA}^+= 0\,, \\
& \hat F_{TAd}- \hat F_{AAd}-\frac 2N F_{TT}^++F_{ST}^+-F_{SA}^-+\frac 2N F_{TA}^-= 0\,, \\
& \hat H_{TAd}- \hat H_{AAd}+\frac 2N H_{TT}^+-H_{ST}^++H_{SA}^--\frac 2N H_{TA}^-= 0\,.
\label{sonsump2b}
\end{split}\ee
Reflection positivity fixes the signs in both the $s$- and $t$-channel for $\smash{\langle \phi_a^i \phi_b^{\bar j,\dagger} \phi_c^k \phi_d^{\bar l,\dagger}\rangle}$. By interchanging the coordinates of the last two fields in the former four-point function and $c \leftrightarrow d$, this then also fixes the signs in the $s$-channel (the channel for which the $\phi \phi^\dagger$ OPE is used) for $\langle \phi_a^i \phi_b^{\bar j,\dagger} \phi_c^{\bar k,\dagger} \phi_d^l\rangle$. The signs in the $t$-channel for the latter four-point function are in turn fixed by reflection positivity. The values of the OPE coefficients in the free-theory limit $d\rightarrow 1$ read
\be\begin{split}
&\lambda^{TAd_+}_l =\lambda^{TAd_-}_l=\lambda^{AAd_+}_l =\lambda^{AAd_-}_l=\lambda^{TT_+}_l =\lambda^{TA_-}_l=\lambda^{AT_-}_l =\lambda^{AA_+}_l=  \frac 12 \lambda_l^{\rm free}\,,  \\
&\lambda^{SAd_+}_l =\lambda^{SAd_-}_l=\lambda^{ST_+}_l =\lambda^{SA_-}_l=\frac{1}{\sqrt{2N}}  \lambda_l^{\rm free}\,,  \\
&\lambda^{TS_+}_l = \lambda^{TS_-}_l = \lambda^{AS_+}_l = \lambda^{AS_-}_l = \frac{1}{2\sqrt{M}}  \lambda_l^{\rm free}\,,  \\
&\lambda^{SS_+}_l= \lambda^{SS_-}_l= \frac 1{\sqrt{2MN}} \lambda_l^{\rm free} \,,
\label{lambdasonsum}
\end{split}\ee
where $\lambda_l^{\rm free}$ is given in eq.~(\ref{Scalar1}) and $l$ is even or odd depending on the representation.
Consistency with the free-theory limit provides a further check on various signs appearing in eqs.~(\ref{sonsump1}) -- (\ref{sonsump2b}).
As a further consistency check, we have verified that eqs.~(\ref{sonsump1}) -- (\ref{sonsump2b}) reduce to eqs.~(\ref{sonsom}) when $\smash{\text{SO}(N)\times\text{SU}(M)\subset\text{SO}(N)\times\text{SO}(2M)}$.

\end{document}